\documentclass{aa}
\usepackage{color}
\usepackage{graphicx}
\usepackage{txfonts}
\usepackage[colorlinks, citecolor=blue]{hyperref}
\usepackage{mathtools}
\usepackage{ulem}
\usepackage{longtable}
\usepackage{lscape}
\usepackage{rotating}
\usepackage{multirow,multicol}

\newcommand{\tE}{\theta_{\text{\scalebox{.8}{E}}}}
\newcommand{\xG}{x_{\text{\scalebox{.8}{G}}}}
\newcommand{\yG}{y_{\text{\scalebox{.8}{G}}}}

\begin{document}
   \title{Gaia GraL II -- {\it Gaia} DR2 Gravitational Lens Systems: \\The Known Multiply Imaged Quasars 
   \thanks{Table of lenses (confirmed and candidates, detected or not in \textit{Gaia} DR2) is only available in electronic form at the CDS via anonymous ftp to cdsarc.u-strasbg.fr (130.79.128.5) or via http://cdsweb.u-strasbg.fr/cgi-bin/gcat?J/A+A/}
}
\author{
C. Ducourant\inst{1},
O. Wertz\inst{2},
A. Krone-Martins\inst{3},
R. Teixeira\inst{4},
J.-F. Le Campion\inst{1},
L. Galluccio\inst{5},
J. Kl{\"u}ter\inst{6},
L. Delchambre\inst{7},
J. Surdej\inst{7}, 
F. Mignard\inst{5},
J. Wambsganss\inst{6},
U. Bastian\inst{6},
M.J. Graham\inst{8},
S.G. Djorgovski\inst{8},
E. Slezak\inst{5}
}
  \institute{
          Laboratoire d'Astrophysique de Bordeaux, Univ. Bordeaux, CNRS, B18N, all{\'e}e Geoffroy Saint-Hilaire, 33615 Pessac, France
          \email{christine.ducourant@u-bordeaux.fr}
         \and
         Argelander-Institut f\"{ u}r Astronomie, Universit\"{ a}t Bonn,  Auf dem H\"{ u}gel 71, 53121 Bonn, Germany
         \and
  		 CENTRA, Faculdade de Ci\^encias, Universidade de Lisboa, Ed. C8, Campo Grande, 1749-016 Lisboa, Portugal
         \and
         Instituto de Astronomia, Geof\'isica e Ci\^encias Atmosf\'ericas, Universidade de S\~{a}o Paulo, Rua do Mat\~{a}o, 1226, Cidade Universit\'aria, 05508-		900 S\~{a}o Paulo, SP, Brazil
         \and         
         Universit\'{e} C\^{o}te d'Azur, Observatoire de la C\^{o}te d'Azur, CNRS, Laboratoire Lagrange, Boulevard de l'Observatoire, CS 34229, 06304 Nice, France
         \and
        Zentrum f\"{u}r Astronomie der Universit\"{a}t Heidelberg, Astronomisches Rechen-Institut, M\"{o}nchhofstr. 12-14, 69120 Heidelberg,\\Germany
         \and
         Institut d'Astrophysique et de G\'{e}ophysique, Universit\'{e} de Li\`{e}ge, 19c, All\'{e}e du 6 Ao\^{u}t, B-4000 Li\`{e}ge, Belgium
         \and
         California Institute of Technology, 1200 E. California Blvd, Pasadena, CA 91125, USA}

   \date{Received May ---, 2018; accepted ---, ---}

  \abstract
   {Thanks to its spatial resolution the ESA/{\it Gaia} space mission offers a unique opportunity to discover new multiply-imaged quasars and to study the already known lensed systems at sub-milliarcsecond astrometric precisions.}
   {In this paper, we address the detection of the known multiply-imaged quasars from the {\it Gaia} Data Release 2 and determine the astrometric and photometric properties of the individually detected images found in the {\it Gaia} DR2 catalogue.} 
   {We have compiled an exhaustive list of quasar gravitational lenses from the literature  to search for counterparts in the {\it Gaia} Data Release 2. We then analyze the astrometric and photometric properties of these {\it Gaia}'s detections. To highlight the tremendous potential of {\it Gaia} at the sub-milliarcsecond level we finally perform a simple Bayesian modeling of the well-known gravitational lens system HE0435-1223,  using {\it Gaia} Data Release 2 and HST astrometry.}
   {From 478 known multiply imaged quasars, 200 have at least one  
   image found in the {\it Gaia} Data Release 2. Among the 41 known quadruply-imaged quasars of the list, 26 have at least one image  in the {\it Gaia} Data Release 2, 12 of which are fully detected (2MASX J01471020+4630433, HE 0435-1223, SDSS1004+4112, PG1115+080, RXJ1131-1231, 2MASS J11344050-2103230, 2MASS J13102005-1714579, B1422+231, J1606-2333, J1721+8842, WFI2033-4723, WGD2038-4008), 6 have three counterparts, 7 have two and 1 has only one. As expected, the modeling of HE0435-1223 shows that the model parameters are significantly better constrained when using {\it Gaia} astrometry compared to HST astrometry, in particular the relative positions of the background quasar source and the centroid of the deflector. The {\it Gaia} sub-milliarcsecond astrometry also significantly reduces the parameter correlations.}
   {Besides providing an up-to-date list of multiply imaged quasars and their detection in the {\it Gaia} DR2, this paper shows that more complex modeling scenarios will certainly benefit from {\it Gaia} sub-milliarcsecond astrometry.}
   

\titlerunning{Gaia GraL II -- The Known Lensed Quasars in {\it Gaia} DR2}
\authorrunning{C. Ducourant et al.}
\maketitle
\section{Introduction}
The ESA/{\it Gaia} space mission \citep{2016A&A...595A...1G} constitutes an exceptional opportunity to characterize and to discover multiply-imaged quasars, although this was not put forth as one of the science objectives in the mission proposal.  With a spatial resolution of $\sim$0.18" {\it Gaia} is roughly comparable to  HST for this particular feature \citep[e.g. ][]{2011Bellini}. However {\it Gaia} being a scanning mission is unique in providing an all-sky coverage with that angular resolution. Thus, by the final {\it Gaia} Data Release (DR),  a whole population of such multiply-imaged quasars would be revealed, providing an all-sky, and the first of this kind, survey of multiply imaged quasars with well understood source detection biases \citep[e.g.][]{2015A&A...576A..74D, 2017A&A...599A..50A, 2018arXiv180409375A}. 

Several programs dedicated to systematic searches for lenses in large astronomical surveys such as SDSS, WISE, DES, PanSTARRS, among others, have been developed in recent years \citep[e.g.][]{2016More, 2017Lin, 2017More, 2018Ostrovski}, and many of them rely on supervised machine learning algorithms trained on simulations to handle the large volume of imaging data \citep[e.g.][]{2017MNRAS.472.1129P, 2017Perreault, 2017Hartley, 2018Pourrahmani, 2018Lanusse}. Naturally, even if {\it Gaia} does not provide images of the observed sources, contrary to the previously mentioned surveys, its high angular resolution over the entire sky is a major asset to contribute to the discovery and the study of multiply-imaged quasars.

\cite{2016Finet} have investigated the potential of Gaia for gravitational lensing and compared it to the detectability with seeing-limited observations for the same limiting magnitude (G = 20). They expect at maximum  about $\sim$1600 multiply imaged quasars with an angular separation large enough to be resolved from the ground in an optimal seeing scenario, while scarcely $\sim80$ would be composed by more than two images. However, they predict that detections from space are much more encouraging, raising the number of multiply imaged quasars detectable by a Gaia-like survey to $\sim$ 2900, thanks to the improved resolving power alone.

The first Gaia Data Release \citep[DR1;][]{2016A&A...595A...2G}, besides providing the best available two-parameter astrometry  (positions only) at the epoch of its publication, did not reach the effective angular resolution necessary to include most of the multiply-imaged quasars. This happened due to data processing issues and final astrometric quality reasons \citep[][]{2016Fabricius, 2017A&A...599A..50A}. Yet, several multiply-imaged quasars discovered from other large surveys such as Pan-STARRS, DES, SDSS-III BOSS, HSCS or VST-ATLAS, were subsequently identified with at least one {\it Gaia} DR1 detection \citep[e.g.][]{2015Agnello, 2017Agnellob, 2017Lemon, 2018Agnello}. 

The {\it Gaia} Data Release 2 \citep[DR2;][]{2018Gaia} on the other hand, starts to reach effective angular resolutions that are capable of resolving more multiply-imaged quasars (expected with typical separations below 1"). The {\it Gaia} DR2 effective resolution reaches $\sim 0.4$", with completeness for separations larger than $\sim 2.2$" \citep{2018Gaia}. However this  resolution  applies strictly only to astrometry and G band photometry; color data are available for objects with separations down to $\sim 2$", but its completeness reaches $\sim 3.5$" \citep{2018Gaia}. Even if it is still far from the ultimate resolving power of the {\it Gaia} instrument, the {\it Gaia} DR2 is a significant advance over DR1, as beyond its much improved effective angular resolution, it contains five-parameters astrometric data (positions, proper-motions, parallax) and also color information for most objects. This simplifies significantly the extraction of genuine extragalactic sources from the galactic stellar contaminants. Only the faintest or most problematic objects are characterized by just a two-parameters solution in this data release, which is unfortunately the case for several multiply imaged quasars since their magnitudes are often close to the {\it Gaia} limiting sensitivity at (G $\sim$ 20.7).  

As part of a larger effort to discover and study multiply-imaged quasar candidates from the various Gaia Data Releases, our group first searched for new lensed systems around known or candidate quasars, enabling the discovery of highly probable multiply-imaged quasar candidates for the first time from {\it Gaia} data alone \citep[{\it Gaia} GraL Paper I;][]{2018arXiv180411051K}. In the present work, we report our findings regarding the identification of known gravitationally lensed quasars in {\it Gaia} DR2. 
We analyze the statistical astrometric properties of the detected lensed images and provide improved relative astrometry for them.
We also derive soft astrometric filters that will be applied, as part of a global blind search ({\it Gaia} GraL paper III; Delchambre et al. in prep), to differentiate foreground stars from extragalactic objects without rejecting the faint components of known lensed systems.
To illustrate how the exquisite optical astrometry of {\it Gaia} at the sub-milliarcsecond level 
may help to better constrain the lenses, we perform a simple modeling of the quadruple lens HE0435-1223 in a Bayesian framework, both using {\it Gaia} and HST astrometry, for comparison purposes. 
 
The paper is organized as follows: In Sect.~\ref{Known} we describe the construction of our  list of gravitationally lensed quasars and candidates from published data. Sect.~\ref{GDR2} presents the matching statistics of this list of known systems with the {\it Gaia} DR2. Sect.~\ref{astrometry} presents the astrometric properties of the {\it Gaia} DR2 data for the known systems. A simple modeling within the Bayesian framework of a known lens using {\it Gaia} DR2 astrometry is described and discussed in Sect.~\ref{Model}. Finally, we summarize our findings in Sect.~\ref{Ccl}.

\section{Compiled list of gravitationally lensed quasars}\label{Known}
We have attempted to compile an as-complete-as-possible list  of known gravitationally lensed quasars published in the literature prior to the {\it Gaia} DR2, including some recent candidates that are not yet spectroscopically confirmed. The major source of known gravitational lenses included in our list is the CASTLES (CfA-Arizona Space Telescope LEns Survey of gravitational lenses) site \citep{1999Castles}\footnote{https://www.cfa.harvard.edu/castles/} providing information for about 100 lenses, most of them observed with HST. Another important single source of known multiply imaged quasars is the SDSS Quasar Lens Search site (SQLS)\footnote{http://www-utap.phys.s.u-tokyo.ac.jp/~sdss/sqls/lens.html}, aimed to discover lensed quasars from the large homogeneous data of the Sloan Digital Sky Survey (SDSS) providing data on 49 additional lensed systems.  We also included several quasar systems from the Master Lens Database \citep{2012Moustakas}\footnote{http://admin.masterlens.org/index.php?}, a community-supported compilation of all discovered strong gravitational lenses. Finally we complemented our list with recent and more scattered discoveries from the literature. This list is being kept up-to-date, and  will be maintained at least until the final {\it Gaia} Data Release. For the sake of completeness, we also included in this list the candidates with indication in the literature of just one image (usually spectroscopic candidates) expecting from the exceptional resolving power of \textit{Gaia} that it may resolve some of them into multiple images in one of its Data Releases. 

Our resulting list of published lensed or lens-candidate quasars contains 478 systems (234 confirmed systems and 244 lensed quasar candidates). This list is only available in electronic form at the CDS, including access through Virtual Observatory ready tools, it comprises  lens identifiers, references and the Gaia astrometry and photometry when a match was found in the DR2. The summarized statistical properties of our list in terms of number of systems with 1, 2, 3 and 4 and more images and status are given in Table\ref{baseAndDR2}.

\section{Gravitationally lensed quasars in Gaia DR2}\label{GDR2}
We extracted sources from the {\it Gaia} Data Release 2 within a radius of 10" around each source of our compiled list of known gravitationally lensed quasars using ADQL and the {\it Gaia} archive facility at ESAC \citep{2017A&C....21...22S}. We obtained the positions ($\alpha, \delta$), parallaxes ($\varpi$), proper-motion components ($\mu_\alpha, \mu_\delta$) and fluxes in the $G$, $G_{\mathrm{BP}}$ and $G_{\mathrm{RP}}$ pass-bands \citep{GaiaDR2Photometry} along with their respective uncertainties. 

For each individual image of each system, we performed a positional cross-match within a maximum angular separation of 0.5" between the astrometry found in the literature and the {\it Gaia} DR2. We visually inspected all systems one by one, by comparing the {\it Gaia} DR2 detections to the system discovery papers and/or archival images from Aladin \citep{2000A&AS..143...33B, 2014ASPC..485..277B}. 

\begin{figure*}[!htp]
\begin{center}
\includegraphics[trim=1.8cm 1cm 0.2cm 1.4cm, clip, width=1\textwidth]{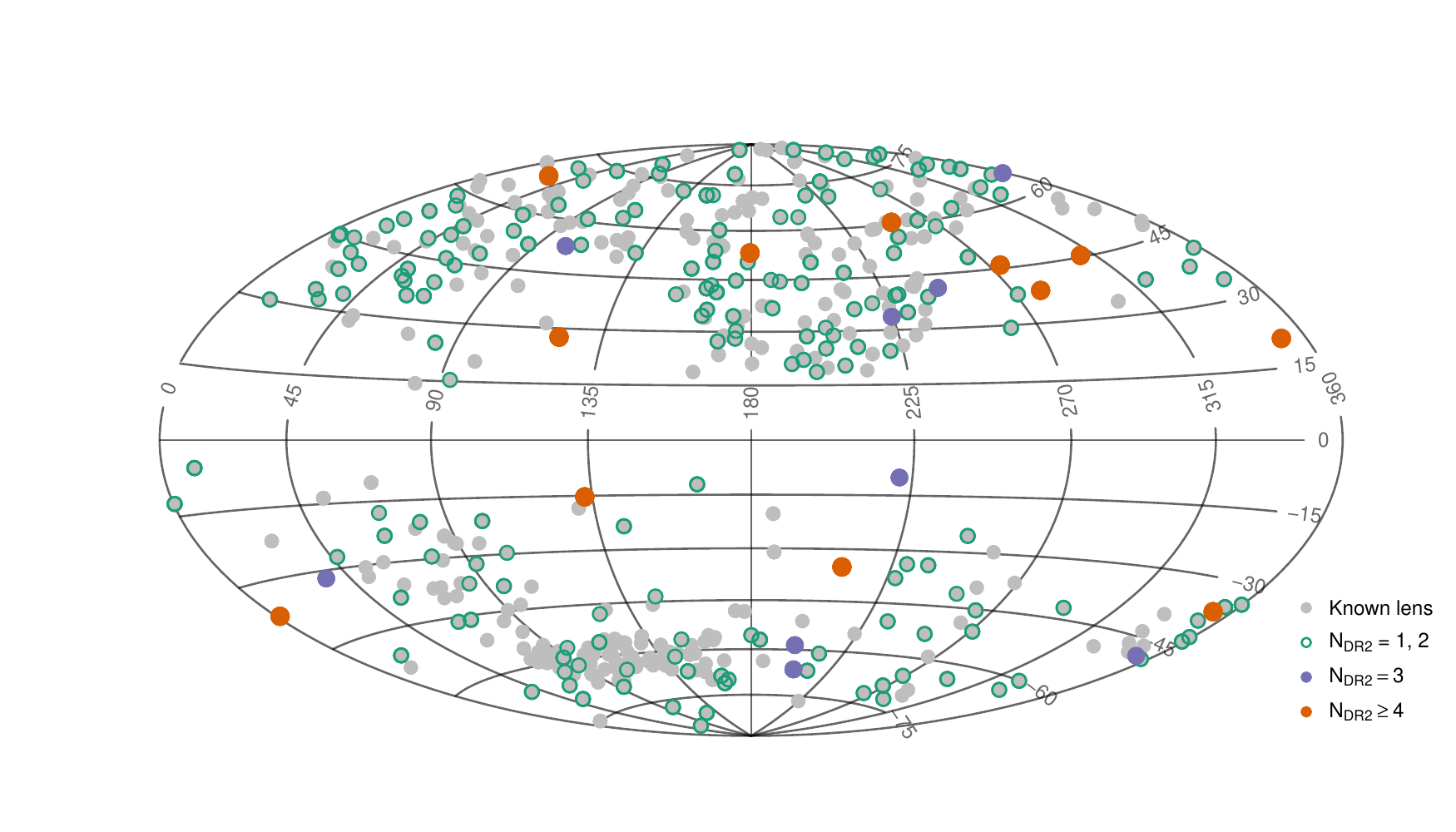}
\caption{All-sky chart in galactic coordinates with the galactic anti-center in the middle. The known multiply-imaged quasars are indicated in gray. The systems presenting one or two counterparts in {\it Gaia} DR2 are surrounded by a green open circle. The systems with three {\it Gaia} DR2 detections are indicated with purple filled circles, while the systems with four or more detections are indicated with orange filled circles.
\label{sky}}
\end{center}
\end{figure*}

 Of the 478 gravitational lens systems (including candidates), 200 have at least one image matched with a \textit{Gaia} DR2 source. The overall detection statistics of known systems 
that result from our examination is given in Table~\ref{baseAndDR2}. An all-sky chart in galactic coordinates of the known lenses is shown in Fig.~\ref{sky} along with a specification of the {\it Gaia} detection.  

In Table \ref{knownlensDR2_34} we present an extract of our list containing the lenses with three and more images for which at least one match was found in the \textit{Gaia} DR2. The complete list of lenses and detection is available in electronic form only at the CDS.

Of the 41 known systems with four images (or more), 26 have at least one image detected in \textit{Gaia} DR2. Within this group, one system has just one image detected, 7 have two images, 6 have three images and 12 are fully detected with four image seen in the \textit{Gaia} data around the target direction. In Fig.~\ref{charts}, we provide charts for the 12 systems with four detections with the Gaia DR2 positions referred to the A image (the brightest image in the system discovery passband) together with flux ratios. Of those which are fully detected, only five  are characterized by sub-milliarcsec astrometry, and are reported in Table \ref{relative_astrometry}. The fainter image (in $G$ band) of the others is detected but poorly constrained.

\begin{table}[!htp]
\centering
\caption{Statistics of the known multiply imaged quasars present in our reference list (col. 2) and of the corresponding detected  systems in the {\it Gaia} Data Release 2 (col. 3). A lensed system is considered to be detected in the {\it Gaia} DR2 if at least one of its images is detected.  Numbers in parentheses correspond to the gravitationally lensed quasar candidates not spectroscopically confirmed yet.}
\label{baseAndDR2}
\begin{tabular}{lrr}
\hline
Lensed	 &  Number of  	  & Number detected \\ 
images	 &	known lenses  & in {\it Gaia} DR2\\	
\hline
1 	     &  55 + (213) = 268      & 11 + (17) =  28\\
2 	     & 133 + (28)  = 161      &112 + (28) = 140\\
3 	     &   6 + (2)   =   8	  &  4 + (2)  =   6\\
4+ 	     &  40 + (1)   =  41	  & 25 + (1)  =  26\\
\hline
Total    &  234+(244) = 478		  &152 +(48) =  200\\
\hline
\end{tabular}
\end{table}

\begin{figure*}[!htp]
\begin{center}
\includegraphics[width=0.32\textwidth]{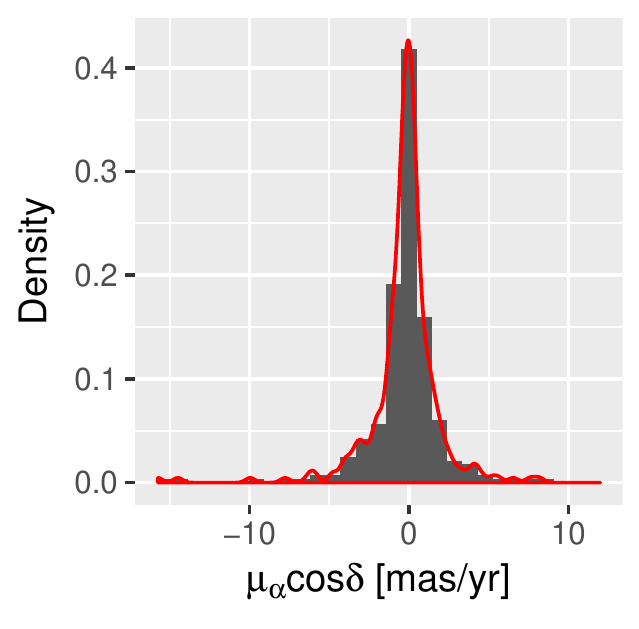}
\includegraphics[width=0.32\textwidth]{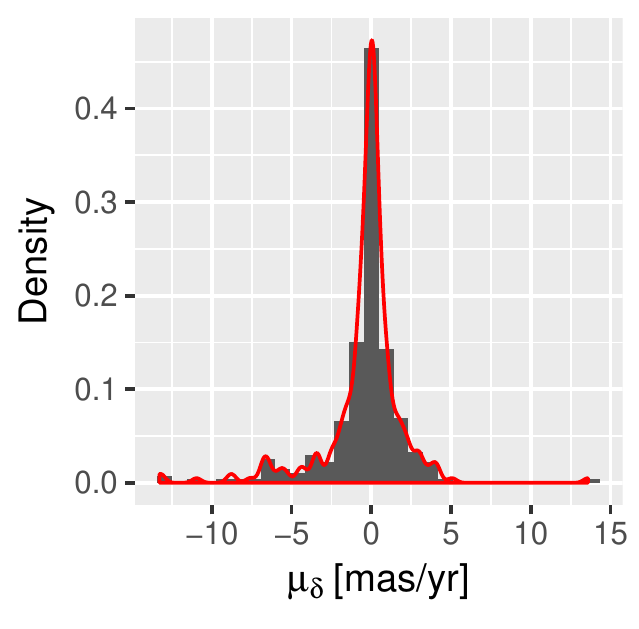}
\includegraphics[width=0.32\textwidth]{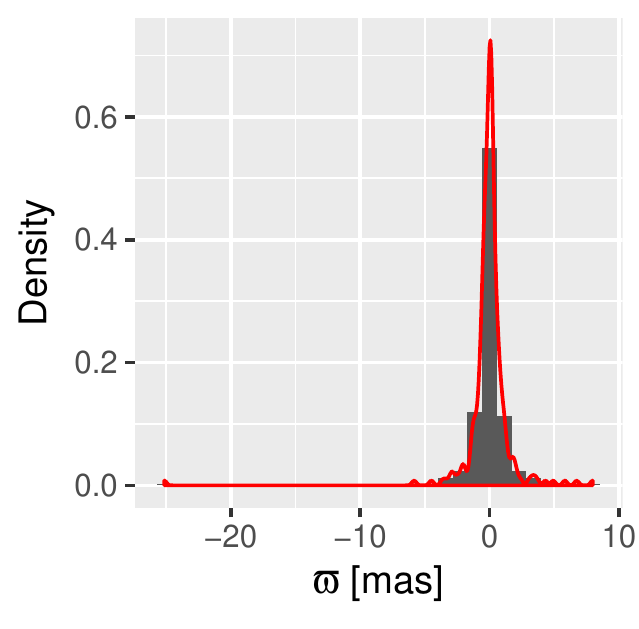}
\includegraphics[width=0.32\textwidth]{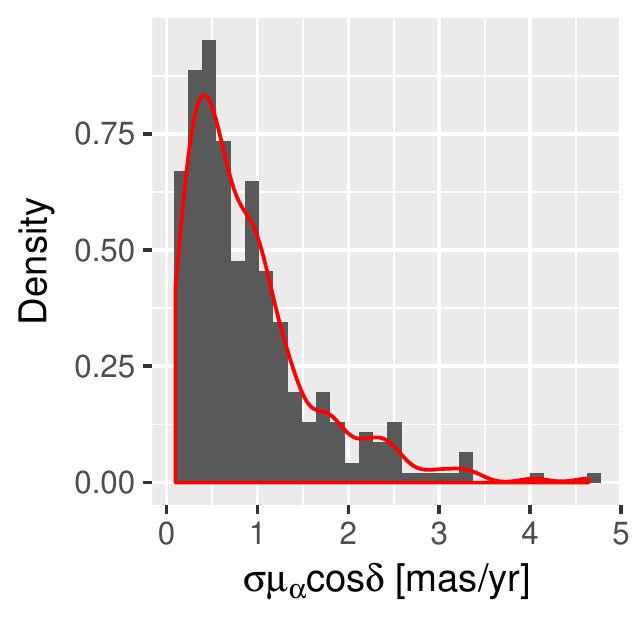}
\includegraphics[width=0.32\textwidth]{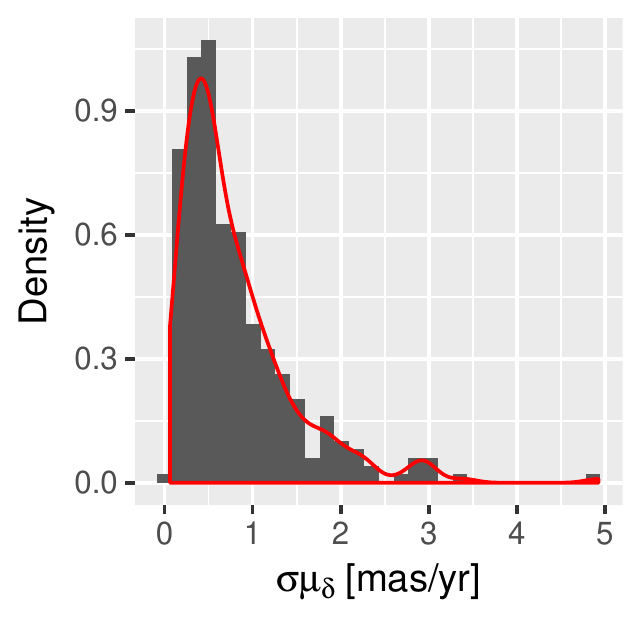}
\includegraphics[width=0.32\textwidth]{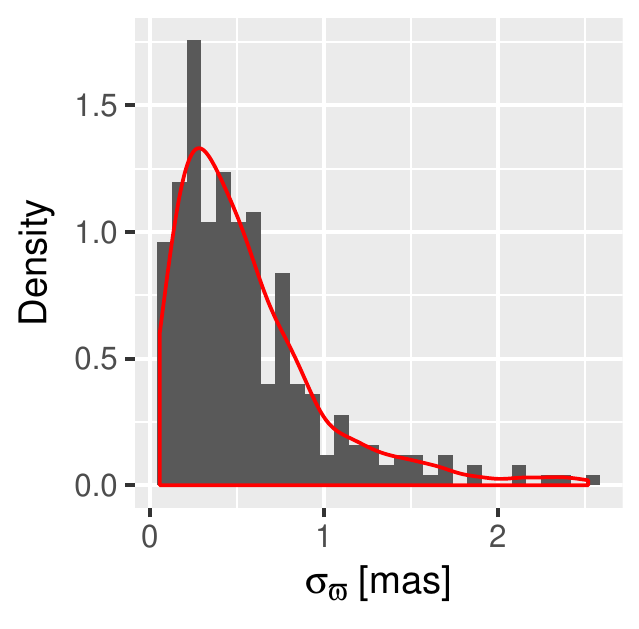}
\caption{Distributions of the astrometric parameters and their uncertainties for all the  {\it Gaia} DR2 counterparts of the individual images of known multiply-imaged quasars with five parameter astrometric solutions.%
\label{astrometry_lenses}}
\end{center}
\end{figure*}

\section{Astrometric properties of the Gaia DR2 gravitationally lensed quasars}\label{astrometry}
The images of a quasar produced by strong gravitational lensing  are peculiar since they are not independent astronomical sources but multiple images of the same source, possibly with part of the host galaxy visible as small segments of an arc. Accordingly, they can produce some particular astrometric signatures in the {\it Gaia} DR2 solution, that could be helpful to discover further lensed quasar systems in the {\it Gaia} data.

Of the 382 individual images found in the {\it Gaia} DR2 coming from the 152 {\it confirmed} gravitationally lensed quasars with two or more images, 65 have a \textit{Gaia}  2-parameter astrometric solution, i.e. right ascension and declination only \citep[see][for a description of the {\it Gaia} DR2 astrometric solution selection]{2018arXiv180409366L}. The other 317 images have complete 5-parameter solutions ($\alpha, \delta, \mu_{\alpha}cos(\delta), \mu_\delta, \varpi$). We investigated the statistical properties of the {\it Gaia} DR2 parameters of the multiple images of the known gravitationally lensed quasars, with results shown in Fig.\ref{astrometry_lenses}. This figure shows that parallaxes and proper motions of the images resulting from lensing occasionally reach large values for sources expected to have neither parallax nor motion.  However, this results from the fact that these images are rather faint, at the limit of the {\it Gaia} DR2 sensitivity where the uncertainties from random noise are also large. In addition it may  be also the case that some sources/images are embedded in extended and diffuse structures. 


In a search for multiply imaged quasars in the {\it Gaia} DR2, applying a straight  astrometric filter aiming at excluding stars from the deviation from zero  parallaxes and proper motions weighted by the expected uncertainties, would likely also exclude a large number of  images of lenses. So, based on the distribution of these parameters for the known lenses, we established the following softer astrometric cuts, that at the expense of a certain level of stellar contamination, avoid the rejection of genuine lens systems or of one or several images within a system.


{\it Gaia DR2} sources that should be accepted to such a search would likely comply with $\varpi$ -3 $\sigma_\varpi$ < 4 mas and 
$|\,\mu\,| - 3 \sigma_{\mu}$ < 4 mas/yr). We note here that $\mu$ stands for $\mu_{\alpha}cos(\delta)$ and $\mu_{\delta}$.
 
Indeed, we also adopted these soft filters in {\it Gaia} GraL Paper I \citep{2018arXiv180411051K}, where we presented the first ever discoveries of quadruply imaged quasar candidates from data of an astrometric space mission. These statistical astrometric properties derived from \textit{Gaia} measurements are also being used in a large, machine learning based, systematic blind-search for lenses in {\it Gaia} DR2 ({\it Gaia} GraL Paper III; Delchambre et al., in prep).

\section{Gravitational lens modeling with sub-mas astrometry}\label{Model}

\begin{table}
\centering
\caption{\label{relative_astrometry}Relative astrometry for five known quadruply imaged quasars  fully detected in the {\it Gaia} DR2. The image references have been chosen to match those reported either in https://www.cfa.harvard.edu/castles/ or in their reference papers. They are not necessarily the brightest images in the \textit{Gaia} G-band.} 
\begin{tabular}{lrr}
\hline
Identifier			& $\Delta\alpha\cos(\delta)$ (mas) & $\Delta\delta$ (mas) \\	
\hline
\hline
HE0435-1223 &                           & \\
A			& $0.0 \pm 0.16$			& $0.0 \pm 0.14$ \\
B			& $-1476.56 \pm 0.19$		& $552.94 \pm 0.16$ \\
C			& $-2466.27 \pm 0.21$		& $-603.05 \pm 0.16$ \\
D			& $-938.66 \pm 0.30$		& $-1614.43 \pm 0.25$ \\
\hline
SDSS1004+4112 & 						& \\
A 			& $0.00 \pm 0.35$ 		    & $0.00 \pm 0.52$ \\
B 			& $1315.29 \pm 0.36$ 		& $3531.57 \pm 0.49$ \\
C 			& $-11039.10 \pm 0.47$ 		& $-4494.69 \pm 0.68$ \\
D 			& $-8403.23 \pm 1.21$ 		& $9701.47 \pm 1.50$ \\
\hline
RXJ1131-1231& 							& \\
A 			& $588.68 \pm 0.36$ 		& $1118.89 \pm 0.23$ \\
B 			& $617.52 \pm 0.39$ 		& $2305.97 \pm 0.25$ \\
C 			& $0.0 \pm 0.47$ 			& $0.0 \pm 0.30$ \\
D 			& $-2522.23 \pm 1.60$ 		& $1993.80 \pm 0.80$ \\
\hline
2MASS J1134-2103		&				& \\
A 			& $0.0 \pm 0.11$ 			& $0.0 \pm 0.07$ \\
B 			& $729.32 \pm 0.11$ 		& $1755.49 \pm 0.07$ \\
C 			& $-1947.39 \pm 0.11$ 		& $-772.70 \pm 0.07$ \\
D 			& $-1247.12 \pm 0.28$ 		& $1366.93 \pm 0.20$ \\
\hline
WFI2033-4723& 							& \\
A1 			& $-2196.54 \pm 0.33$ 		& $1261.42 \pm 0.34$ \\
A2 			& $-1483.18 \pm 0.26$ 		& $1375.75 \pm 0.30$ \\
B 			& $0.0 \pm 0.27$ 			& $0.0 \pm 0.24$ \\
C 			& $-2113.84 \pm 0.33$ 		& $-277.84 \pm 0.32$ \\
\hline
\end{tabular}
\end{table}


\begin{figure*}
	\centering
	\includegraphics[width=0.95\textwidth]{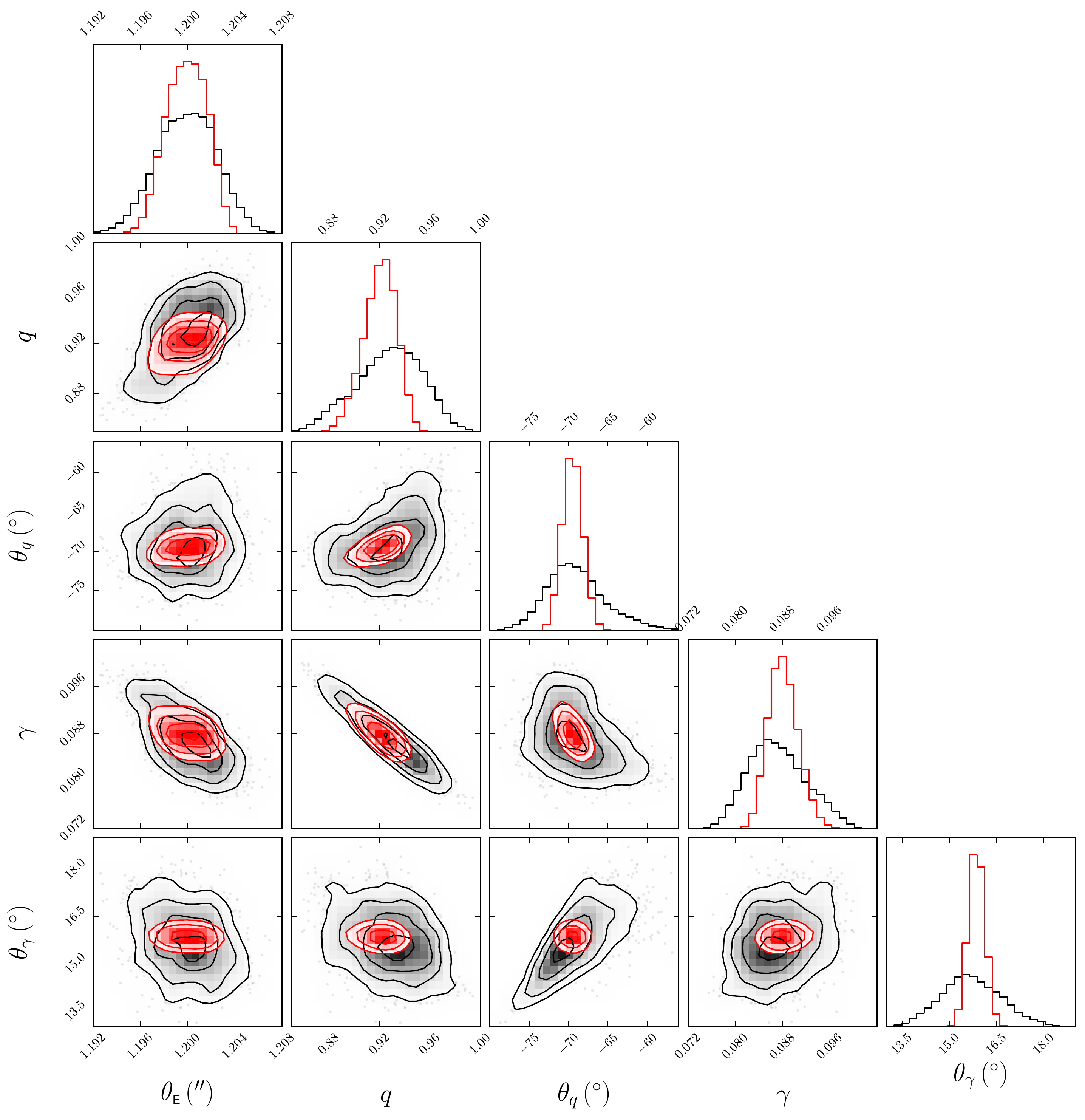}
     	\caption{	Results of the MCMC simulations for HE0435-1223, displayed as a corner plot for the five model parameters. 
			The diagonal panels illustrate the posterior PDFs while the off-axis panels illustrate the correlation between the 
			parameters. We show the results obtained from {\it Gaia}'s data with shaded red contours and red histograms and  with HST data with shaded black contours and black histograms. The three inner contours represent the $68.3\%$, $95.4\%$, and $99.7\%$ confidence intervals.
		    }
       	\label{figure:HE0435_contour_nie}       
\end{figure*}

\begin{figure*}
	\centering
    \includegraphics[width=0.95\textwidth]{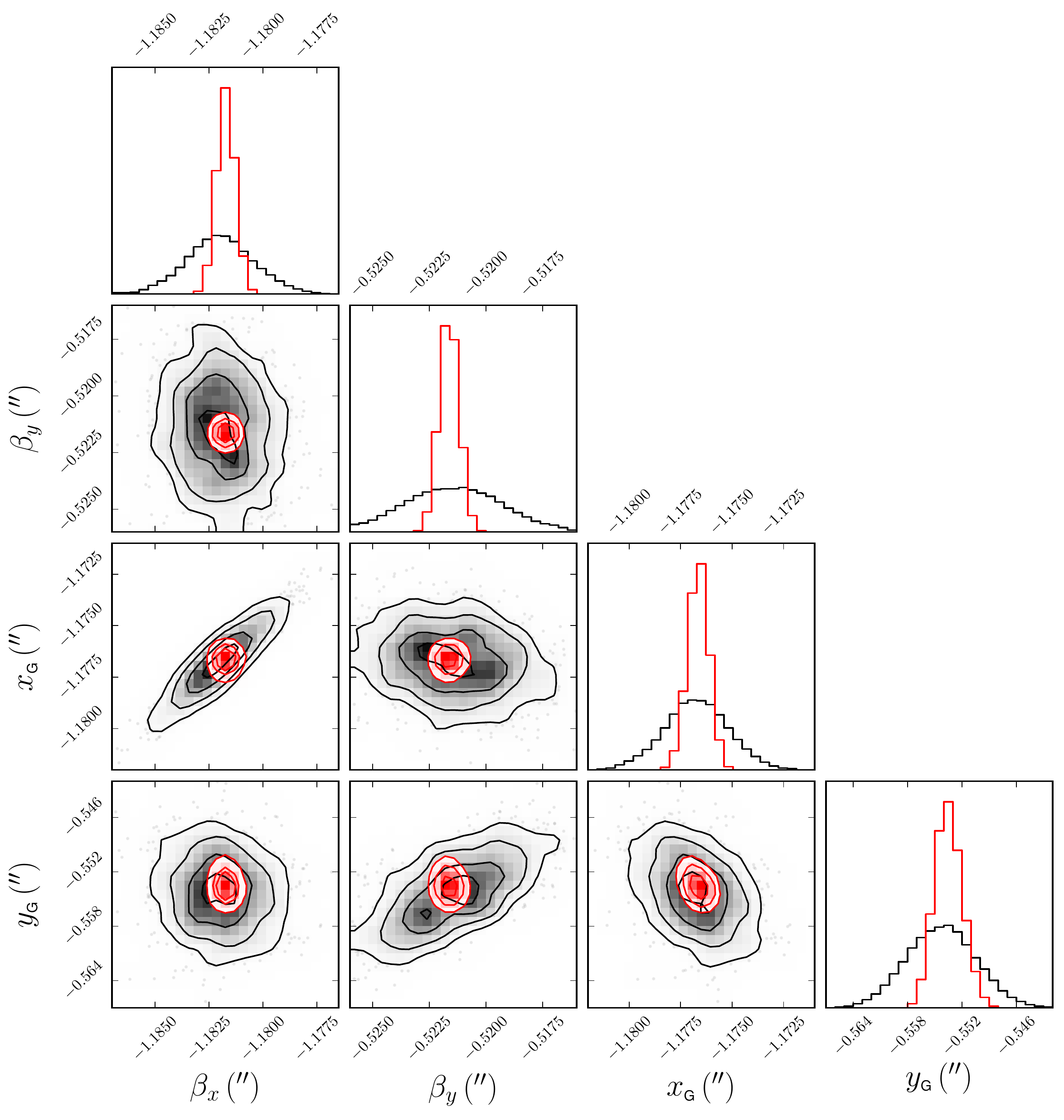}
    \caption{Results of the MCMC simulations for HE0435-1223, displayed as a corner plot for the source and deflector positions. The diagonal panels illustrate the posterior PDFs while the off-axis panels illustrate the correlation between the parameters. We show the results obtained from {\it Gaia}'s data with shaded red contours and red histograms and  with HST data with shaded black contours and black histograms. The three inner contours represent the $68.3\%$, $95.4\%$, and $99.7\%$ confidence intervals.
		     }
    \label{figure:HE0435_contour_srcgal}       
\end{figure*}

Gravitational lensing provides an efficient tool to explore various aspects of our universe and several of its components. 
In the strong regime, the inference of physically meaningful quantities from observational data usually requires the accurate modeling of the gravitational potential of the  deflector. For example, the ability of modern time 
delay cosmography to infer the Hubble constant $H_0$ with a competitive precision relies significantly on its capacity in 
dealing with families of degeneracies existing between different plausible lens mass profiles \citep[][]{Saha_degeneracies_2000, Wucknitz_TD_2002, Liesenborgs_2012, SPT_SS13}. 
To probe the deflector mass distribution in the region where multiple images are formed, simple parametrized mass models 
are commonly used whose parameters are fixed by the observational constraints \citep[e.g.][]{Keeton_computational_methods_2001, Keeton_modeling_2010, Lefor_Futamase_2013, Lefor_2014}, 
typically the  lensed quasar image positions, the morphology of extended components, microlensing-free flux ratios, 
and time delays between image pairs. Naturally, a better accuracy in the observed parameters leads to a more reliable model. 
For the five known lenses 
RXJ1131-1231, 
SDSS1004+4112, 
2MASS J1134-2103, 
HE0435-1223, 
and WFI2033-4723, 
the {\it Gaia} DR2 provides quasar image position measurements with an unprecedented precision of a few tenths of a milliarcsecond. 
With an order of magnitude improvement over typical HST astrometric accuracy, these new astrometric data should help 
to better constrain the  lens models. Considering four of the five known quadruply imaged quasars reported in 
Table 2 for which Gaia and HST astrometric data are available, the average of the Gaia astrometric uncertainties affecting 
the equatorial coordinates of the four lensed quasar images is found to be 0.43 mas compared to 3.29 mas using the 
corresponding HST data. This represents a huge gain (by more than a factor 7) in astrometric precision.    

In this section, we illustrate how the improved astrometric accuracy obtained with {\it Gaia} may impact the lens modeling.   To this end, we propose to optimize a smooth model to the observed image positions only, within the Bayesian framework.  The idea consists in simultaneously sampling the posterior Probability Density Functions (PDFs) for all model parameters using a Markov Chain Monte Carlo (MCMC) method, and then comparing the PDFs obtained from Gaia's astrometry with  the ones derived from the astrometry found in the literature. We want to point out that our objective here is not to construct  a set of realistic lens models in the sense that they could be used to perform time delay cosmography. Instead, we focus on how {\it Gaia} astrometric uncertainties may positively affect the goodness of a more complex fit, which would include microlensing-free flux ratios, time delays, non-lensing data related to the main deflector, or even simultaneous reconstruction of the source and deflector surface brightnesses.

We model the main deflector as a singular isothermal ellipsoid \citep[SIE, see e.g.][]{Kormann_SIE_1994} which effectively describes the mass distribution of a massive early-type galaxy in the region where multiple images are formed \citep[][]{Gilman_2017}. An SIE is characterized by five free parameters; the Einstein radius $\tE$, the elliptical axes ratio $q$ and position angle $\theta_q$, and the lens centroid $(\xG,\yG)$ with respect to image A. Since the lensed quasar image positions are generally most sensitive to the local mass distribution, we model the large-scale contributions and possible close line-of-sight galaxy perturbing effects with an external shear term characterized by its absolute value $\gamma$ and position angle $\theta_{\gamma}$. The model is thus kept simple, which limits the number of free parameters and avoids the use of the full multi-plane lensing formalism \citep[][]{Schneider_MST_multiplane_2014, McCully_2014, McCully_2016}. In addition, we consider the position of the point-like source $(\beta_x, \beta_y)$ with respect to image A that is also free to vary during the optimization process, bringing the number of free parameters $n_k$ to nine.

To draw samples from the posterior PDFs, we used \texttt{emcee}\footnote{https://github.com/dfm/emcee} \citep[][]{emcee}, a python package which implements the affine 
invariant ensemble sampler for MCMC proposed by \citet[][]{Goodman_MCMC_2010}. Since we only use the lensed image positions $\boldsymbol{\theta}_{\text{obs}}$
as observational data to fit, the log-likelihood function simply reads
\begin{equation}
	\ln{p(\boldsymbol{\theta}_{\text{obs}}|\boldsymbol{k})} = -\frac{1}{2} \sum_{j=1}^{2N}\left( \frac{\left(\theta_{\text{obs},j} - \theta_{\text{model},j}(\boldsymbol{k})\right)^2}{\sigma_{\text{obs},j}^2} - \ln{\left(\sigma_{\text{obs},j}^2\right)} \right) \, ,
	\label{eq:lnprob}
\end{equation}
where $\boldsymbol{k}$ is the vector of free parameters, $N$ the number of lensed images (hence $2N$ constraints), $\sigma_{\text{obs}}$ 
the astrometric uncertainties, and $\boldsymbol{\theta}_{\text{model}}(\boldsymbol{k})$ the lensed image positions obtained from the 
free parameters $\boldsymbol{k}$ and generated with the python package $\texttt{pySPT}$\footnote{https://github.com/owertz/pySPT} \citep[][]{pySPT_2018}.
To control the sampling, only two hyperparameters need to be tuned: an adjustable scale parameter $a$ and the number $N_{w}$ of walkers. 
The scale parameter $a$ has a direct impact on the acceptance rate of each walker, namely the ratio of accepted to proposed candidates, 
and was set to $a = 2$, following \citet[][]{Goodman_MCMC_2010}.
A walker can be seen as a Metropolis-Hastings chain \citep[see, e.g.,][]{MacKay_2003_InformationTheory} whose associated proposal 
distribution depends on the positions of all the other walkers \citep[][]{emcee}. Prior to run the MCMC, we initialized $N_w = 350$ walkers in a small $n_{k}$-dimensional 
ball of the parameter space around a highly probable solution, formerly obtained using the public lens modeling code \texttt{lensmodel}\footnote{http://physics.rutgers.edu/~keeton/gravlens/2012WS/} \citep[v1.99,][]{Keeton_computational_methods_2001}. Obviously this initial model may not be the most appropriate one and is more likely a local solution in the parameter space. Nevertheless, it constitutes a valid starting point to illustrate our intention.

The analysis has been performed for the five lenses from Table \ref{relative_astrometry} for which HST image position measurements are available, namely 
HE0435-1223, 
SDSS1004+4112, 
RXJ1131-1231, 
2MASS J1134-2103,
and WFI2033-4723.
In Figs. \ref{figure:HE0435_contour_nie} and \ref{figure:HE0435_contour_srcgal}, we illustrate the MCMC results in the form of corner plots for HE0435-1223, which are representative of each of the five quadruply imaged quasars that  we have preliminary modeled. 
The HE0435-1223 image positions measured with the Wide Field Camera 3 \citep[IR/F160W, ][]{HST_WFC3_2002}  mounted on the HST come from \citet[][]{HE0435_HST_Kochanek_2006}, showing astrometric uncertainties between $3$ and $5$ mas.
As expected, all the posterior PDFs obtained from {\it Gaia}'s data show narrower widths than those obtained from HST data, while some of them are slightly shifted.
Thus the use of {\it Gaia}'s astrometry significantly reduces the ranges of valid model parameters around a highly probable solution, as shown in Table \ref{confidence_intervals}. 
 The least sensitive parameter is the Einstein radius, even if it is improved three-folded with respect to HST observations. In particular, the source position is constrained within a $\sigma$-error ellipse of $(\sigma_{\beta_x}, \sigma_{\beta_y}) = (0.1,0.1)$ mas. 
This one order of magnitude improvement indicates that the sub-mas astrometry of {\it Gaia} clearly  helps to better constrain the position of the point-like source as well as the source surface brightness reconstruction as part of a more realistic modeling scenario. 

We also note that the {\it Gaia} DR2 astrometry reduces significantly the resulting correlation structure between the modeled parameters, in comparison with the correlations obtained from the modeling using HST data: the absolute value of the correlation coefficients between $\theta_q$ and $\theta_\gamma$, and  $\theta_q$ and $q$, in Fig.~\ref{figure:HE0435_contour_nie} and between $\beta_y$ and $\yG$, and $\beta_x$ and $\xG$ in Fig.~\ref{figure:HE0435_contour_srcgal}, are clearly reduced thanks to the improved astrometry.


A more advanced version of the lens modeling within the Bayesian framework described in this section will be consistently applied to all the known lenses and to the highly probable lens candidates discovered from the systematic blind-search for lenses in the entire {\it Gaia} DR2 ({\it Gaia} GraL Paper III, Delchambre et al., in prep), and this will be presented in a forthcoming work ({\it Gaia} GraL Paper IV, Wertz et al., in prep).
 

\begin{table}
\centering
\caption{\label{confidence_intervals}The SIEg lens model parameters derived for HE0435-1223. The reported values are medians within $1\sigma$ error bars.} 
\begin{tabular}{lrr}
\hline
Parameters						& HST & {\it Gaia} \\	
\hline
\hline
$\tE$ (") 						& $1.2 \pm 0.003$ 		& $1.2 \pm 0.001$ \\
$q$ 							& $0.93 \pm 0.03$ 		& $0.9210 \pm 0.01$ \\
$\theta_q$ ($^{\circ}$)			& $-69.4 \pm 3.8$ 		& $-69.5 \pm 0.8$ \\
$\gamma$ 						& $0.087 \pm 0.005$		& $0.088 \pm 0.002$ \\
$\theta_{\gamma}$ ($^{\circ}$) 	& $15.7 \pm 0.9$		& $15.9 \pm 0.1$\\
\hline
$\beta_{x}$ (mas)				& $-1182.0 \pm 1.6$ 	& $-1181.7 \pm 0.1$ \\
$\beta_{y}$ (mas)				& $-521.4 \pm 2.1$ 		& $-521.6 \pm 0.1$ \\
$\xG$ (mas)						& $-1176.7 \pm 1.5$ 	& $-1176.6 \pm 0.3$ \\
$\yG$ (mas)						& $-554.2 \pm 3.7$ 		& $-553.6 \pm 1.2$ \\
\hline
\end{tabular}
\end{table}




\section{Conclusions}\label{Ccl}
The availability of high-precision and high-accuracy astrometric data as provided by the ESA/\textit{Gaia} space mission opens a new window to detect and model gravitationally lensed quasar systems with an unprecedented refinement. This is bound to impact on fundamental applications in astronomy that are derived from this phenomena, such as the study of the lensing galaxy populations, distant quasars, dark matter and dark energy properties and consequently the determination of cosmological parameters.  To exploit this new field with \textit{Gaia} data we have set up a collaboration group, the {\it Gaia} GraL team, to systematically analyze the gravitationally lensed quasar content throughout the {\it Gaia} Data Releases. The topics covered include searches for new multiply imaged quasar candidates,  identifications of known lenses in the \textit{Gaia} data,   modeling of the lenses using the outstanding \textit{Gaia}  astrometry and multi-colour photometry, and fostering ground-based follow-up for final confirmation.

In this paper we explain how we first generated an up-to-date list of known gravitationally lensed quasars, including lensed quasars too faint to be observed by {\it Gaia}.  The {\it Gaia} GraL list of known gravitationally lensed quasars will be kept up-to-date with respect to the astronomical literature at least until the final {\it Gaia} Data Release. Each {\it Gaia} Data Release will be analyzed to verify the detection of known gravitational lenses. 

Then  we provide here the first ever sub-milliarcsecond astrometric data for hundreds of known gravitationally lensed quasars. The search is based on the aforementioned list matched to the {\it Gaia} DR2 astrometric catalogue, the largest and most precise astrometric reference available to date. Our lens results bring almost  one order of magnitude improvement in astrometric precision compared to a typical HST observation. Moreover, even if {\it Gaia} DR2 is still an early Data Release lacking many lensed images, it brings high-precision astrometry complemented with photometric data for most known lensed systems. Thus, it provides a glimpse of the content that will become available in the forthcoming Gaia Data Releases.

Of the 478 presently known  – or candidate – gravitationally lensed quasars, we have found in the {\it Gaia} DR2 at least one counterpart for 200 of them. From these objects, the quadruply-imaged quasars occupy a specially relevant place, as they provide the more stringent physical parameter inferences. There are 41 presently known quads. From these, 25 have been found with at least one entry in {\it Gaia} DR2 and 12 of them are fully detected with all four images. As the images of many of these objects have smaller angular separations than the {\it Gaia} DR2 best angular resolution, we expect however the forthcoming Data Releases to provide information for most of them when the releases gradually reach the  expected {\it Gaia} best spatial resolution. We provide also {\it Gaia} DR2 astrometric and photometric data for all known lenses to date.

Finally, we show that the adoption of high-precision astrometry from {\it Gaia} DR2 to model the well-known lens system HE0435-1223 results in a significant improvement in constraining the lens parameters of a NSIEg model around a highly probable solution, and that it also significantly reduces the parameter correlations, in comparison to standard HST astrometry. Such constraints will certainly be further improved with the increased precision of  {\it Gaia}'s  forthcoming nominal mission Data Releases, expected for 2020 (DR3) and 2022 (DR4),  
and the still to be announced Data Release(s) of the {\it Gaia} mission extension. 

As a final conclusion this work vividly demonstrates the significant impact of high-precision astrometry from {\it Gaia} and future mission concepts as the JASMINE series \citep{Gouda:2011}, GaiaNIR \citep{2016arXiv160907325H}, and Theia \citep{2017arXiv170701348T}, to the study of strong gravitational lensing. This paper also exemplifies the ever wider  impact of the {\it Gaia} satellite, pushing its limits from its original goal of studying the Milky Way galaxy towards  more distant extragalactic sources and associated phenomena.

\begin{acknowledgements}
       AKM acknowledges the support from the Portuguese Funda\c c\~ao para a Ci\^encia e a Tecnologia (FCT) through grants SFRH/BPD/74697/2010, from the Portuguese Strategic Programme UID/FIS/00099/2013 for CENTRA, from the ESA contract AO/1-7836/14/NL/HB and from the Caltech Division of Physics, Mathematics and Astronomy for hosting a research leave during 2017-2018, when this paper was prepared. 
      LD and JS acknowledge support from the ESA PRODEX Programme `Gaia-DPAC QSOs' and from the Belgian Federal Science Policy Office.
      OW acknowledges support from a fellowship for Postdoctoral Researchers by the Alexander von Humboldt Foundation.
      SGD and MJG acknowledge a partial support from the NSF grants AST-1413600 and AST-1518308, and the NASA grant 16-ADAP16-0232.
      We acknowledge partial support from `Actions sur projet INSU-PNGRAM', and from the Brazil-France exchange programmes Funda\c c\~ao de Amparo \`a Pesquisa do Estado de S\~ao Paulo (FAPESP) and Coordena\c c\~ao de Aperfei\c coamento de Pessoal de N\'ivel Superior (CAPES) -- Comit\'e Fran\c cais d'\'Evaluation de la Coop\'eration Universitaire et Scientifique avec le Br\'esil (COFECUB).
      The authors wish to thank C. Spindola Duarte for her help with the source referencing.      
      This work has made use of the computing facilities of the Laboratory of Astroinformatics (IAG/USP, NAT/Unicsul), whose purchase was made possible by the Brazilian agency FAPESP (grant 2009/54006-4) and the INCT-A, and we thank the entire LAi team, specially Carlos Paladini, Ulisses Manzo Castello, and Alex Carciofi for the support.
      	This research has made use of the VizieR catalogue access tool, CDS, Strasbourg, France. The original description of the VizieR service was published in A\&AS 143, 23. This research has made use of ``Aladin sky atlas'' developed at CDS, Strasbourg Observatory, France.
      This work has made use of results from the ESA space mission {\it Gaia}, the data from which were processed by the {\it Gaia} Data Processing and Analysis Consortium (DPAC). Funding for the DPAC has been provided by national institutions, in particular the institutions participating in the {\it Gaia} Multilateral Agreement. The {\it Gaia} mission website is:
http://www.cosmos.esa.int/gaia. Some of the authors are members of the {\it Gaia} Data Processing and Analysis Consortium (DPAC).
\end{acknowledgements}

\bibliographystyle{aa}
\bibliography{bibliography} 
\clearpage
\onecolumn
\longtab{
\begin{landscape}
\scriptsize
\begin{longtable}{llrrrrrrr}
\caption{\label{knownlensDR2_34} List of known gravitationally lensed quasars with 3, 4 or 5 images and with at least one match in the \textit{Gaia} DR2. (1)Name (with (*) see note at the bottom), (2) ref - bibliographic reference (* designates candidates), (3) $N_{im}$ – number of images of the lens in the literature, (4) Gaia SourceId, (5,6) ICRS positions from \textit{Gaia} DR2 at epoch 2015.5, (7,8,9) \textit{Gaia} G, $G_{\mathrm{BP}}$, $G_{\mathrm{RP}}$ magnitudes and standard errors (calculated by CDS).}\\
\hline
Name &ref& $N_{im}$ & Gaia SourceId & Right ascension & Declination &  $G$ & $G_{\mathrm{BP}}$ & $G_{\mathrm{RP}}$ \\ 
     &         &       &           &[$^\circ$]$\pm$[mas]        &[$^\circ$]$\pm$[mas] & [mag] & [mag] & [mag] \\
\hline
\endfirsthead
\caption{continued.}\\
\hline\hline
Name &ref& $N_{im}$ & Gaia SourceId & Right ascension & Declination &  $G$ & $G_{\mathrm{BP}}$ & $G_{\mathrm{RP}}$ \\ 
     &         &       &           &[$^\circ$]$\pm$[mas]        &[$^\circ$]$\pm$[mas] & [mag] & [mag] & [mag] \\
\hline
\endhead
\hline
\endfoot
\endfoot
2MASX J01471020+4630433  B  &  33  & 4 & 350937280925970432  &  26.79193807656 $\pm$ 0.1330 &  46.51214243036 $\pm$ 0.2039 & 16.7405 $\pm$ 0.0031 &                    &                    \\ 
2MASX J01471020+4630433  D  &  33  & 4 & 350937280925971456  &  26.79230269485 $\pm$ 0.2699 &  46.51127175724 $\pm$ 0.3121 & 18.2644 $\pm$ 0.0026 & 18.3850 $\pm$ 0.1694 & 17.5129 $\pm$ 0.0107\\ 
2MASX J01471020+4630433  A  &  33  & 4 & 350937280928094336  &  26.79244106590 $\pm$ 0.4754 &  46.51216827296 $\pm$ 0.4487 & 15.8931 $\pm$ 0.0075 & 15.5900 $\pm$ 0.0548 & 14.8218 $\pm$ 0.0441\\ 
2MASX J01471020+4630433  C  &  33  & 4 & 350937280925970304  &  26.79291224025 $\pm$ 4.7897 &  46.51205497601 $\pm$ 2.3003 & 16.1790 $\pm$ 0.0067 &                    &                    \\ 
\hline
HE0230-2130              B  &  43  & 4 & 5126705515510051584 &  38.13811782066 $\pm$ 0.2498 & -21.29074906026 $\pm$ 0.4626 & 19.3041 $\pm$ 0.0078 &                    &                    \\ 
HE0230-2130              A  &  43  & 4 & 5126705515511054080 &  38.13832537061 $\pm$ 0.1989 & -21.29067749450 $\pm$ 0.3369 & 19.0355 $\pm$ 0.0047 & 18.3755 $\pm$ 0.0226 & 17.7388 $\pm$ 0.0336\\ 
HE0230-2130              C  &  43  & 4 & 5126705515510051712 &  38.13847486587 $\pm$ 0.4449 & -21.29024032275 $\pm$ 0.6378 & 20.0651 $\pm$ 0.0141 &                    &                    \\ 
\hline
WISE 025942.9-163543     C  &  42  & 4 & 5153828508862118912 &  44.92836484851 $\pm$ 0.9512 & -16.59520492683 $\pm$ 0.6912 & 20.1708 $\pm$ 0.0225 &                    &                    \\ 
WISE 025942.9-163543     B  &  42  & 4 & 5153828504567283072 &  44.92869905654 $\pm$ 0.5370 & -16.59511750800 $\pm$ 0.4263 & 19.6393 $\pm$ 0.0131 &                    &                    \\ 
WISE 025942.9-163543     A  &  42  & 4 & 5153828508862119040 &  44.92879162716 $\pm$ 0.5258 & -16.59536150645 $\pm$ 0.4166 & 19.4150 $\pm$ 0.0084 & 18.9059 $\pm$ 0.0573 & 17.6442 $\pm$ 0.0325\\ 
\hline
J0408-5354               C*  &  23  & 3 & 4779903605191703296 &  62.09056074627 $\pm$ 3.5274 & -53.89942733757 $\pm$ 2.5860 & 20.8249 $\pm$ 0.0222 &                    &                    \\ 
J0408-5354               B*  &  23  & 3 & 4779902849277205760 &  62.09142458061 $\pm$ 0.5550 & -53.90025030949 $\pm$ 0.5666 & 20.4514 $\pm$ 0.0101 & 20.2464 $\pm$ 0.0971 & 19.6238 $\pm$ 0.0610\\ 
J0408-5354               A*  &  23  & 3 & 4779903605192400000 &  62.08965552446 $\pm$ 0.3402 & -53.89973595868 $\pm$ 0.3595 & 19.7447 $\pm$ 0.0071 & 19.7080 $\pm$ 0.0480 & 18.7889 $\pm$ 0.0462\\ 
\hline
HE0435-1223              C  &  31  & 4 & 3178020716638059136 &  69.56160351805 $\pm$ 0.1802 & -12.28748166882 $\pm$ 0.1294 & 18.8373 $\pm$ 0.0067 & 18.5996 $\pm$ 0.0952 & 17.8801 $\pm$ 0.0610\\ 
HE0435-1223              B  &  31  & 4 & 3178020716640423680 &  69.56188488232 $\pm$ 0.1534 & -12.28716055854 $\pm$ 0.1249 & 18.8942 $\pm$ 0.0104 &                    &                    \\ 
HE0435-1223              D  &  31  & 4 & 3178020716638059392 &  69.56203780095 $\pm$ 0.2818 & -12.28776263363 $\pm$ 0.2298 & 19.3013 $\pm$ 0.0104 &                    &                    \\ 
HE0435-1223              A  &  31  & 4 & 3178020716638059264 &  69.56230465346 $\pm$ 0.1145 & -12.28731415348 $\pm$ 0.0978 & 18.5754 $\pm$ 0.0060 & 18.0642 $\pm$ 0.0452 & 17.6339 $\pm$ 0.0220\\ 
\hline
J0630-1201               C  &  26  & 3 & 3000185396723852416 &  97.53784635400 $\pm$ 1.6412 & -12.02241686045 $\pm$ 2.1094 & 20.0917 $\pm$ 0.0083 & 19.5704 $\pm$ 0.2020 & 18.4840 $\pm$ 0.0277\\ 
J0630-1201               A  &  26  & 3 & 3000185396729743104 &  97.53799106290 $\pm$ 0.8450 & -12.02225598303 $\pm$ 1.0049 & 19.9856 $\pm$ 0.0067 & 19.5925 $\pm$ 0.0622 & 18.1779 $\pm$ 0.0373\\ 
J0630-1201               B  &  26  & 3 & 3000185396723852544 &  97.53808564125 $\pm$ 0.5711 & -12.02194351373 $\pm$ 0.6697 & 20.0078 $\pm$ 0.0071 & 19.2062 $\pm$ 0.1361 & 18.0411 $\pm$ 0.0467\\ 
\hline
HS0810+2554              A  &  43  & 4 & 682614034415644032  & 123.38030267112 $\pm$ 0.6695 &  25.75086095081 $\pm$ 0.4691 & 16.1001 $\pm$ 0.0062 & 16.1982 $\pm$ 0.0176 & 15.1723 $\pm$ 0.0078\\ 
HS0810+2554              D  &  43  & 4 & 682614034415644160  & 123.38049319374 $\pm$54.2925 &  25.75102468034 $\pm$36.3600 & 18.9633 $\pm$ 0.0116 &                    &                    \\ 
\hline
RXJ0911+0551             D  &  43  & 4 & 580537088586198656  & 137.86423190910 $\pm$ 0.6464 &	5.84851723352 $\pm$ 0.4200 & 19.8927 $\pm$ 0.0085 & 19.8129 $\pm$ 0.0927 & 19.1918 $\pm$ 0.1998\\ 
RXJ0911+0551             C  &  43  & 4 & 580537092879166720  & 137.86506252061 $\pm$48.4513 &	5.84856614310 $\pm$21.7528 & 19.7670 $\pm$ 0.0157 &                    &                    \\ 
RXJ0911+0551             B  &  43  & 4 & 580537092879166848  & 137.86513579392 $\pm$ 0.5169 &	5.84840999370 $\pm$ 0.5057 & 18.7832 $\pm$ 0.0120 & 18.3875 $\pm$ 0.0383 & 17.8027 $\pm$ 0.0263\\ 
\hline
SDSS0924+0219            A  &  43  & 4 & 3844748070752040576 & 141.23256103405 $\pm$ 0.2185 &	2.32371368311 $\pm$ 0.2048 & 18.3811 $\pm$ 0.0141 & 18.3105 $\pm$ 0.0502 & 17.6076 $\pm$ 0.05273\\ 
SDSS0924+0219            B  &  43  & 4 & 3844748075046370688 & 141.23257901195 $\pm$ 0.5972 &	2.32321168663 $\pm$ 0.6246 & 19.9453 $\pm$ 0.0145 &                    &                    \\ 
\hline
SDSS1004+4112            C  &  43  & 5 & 806853174702355072  & 151.14095539984 $\pm$ 0.3976 &  41.20964940678 $\pm$ 0.5731 & 19.9953 $\pm$ 0.0087 & 19.8931 $\pm$ 0.0883 & 19.3598 $\pm$ 0.0514\\ 
SDSS1004+4112            D  &  43  & 5 & 806853174702356352  & 151.14192866068 $\pm$ 1.1881 &  41.21359278404 $\pm$ 1.4523 & 20.5619 $\pm$ 0.0130 & 20.2623 $\pm$ 0.2021 & 19.6826 $\pm$ 0.0808\\ 
SDSS1004+4112            A  &  43  & 5 & 806853178999388928  & 151.14503143567 $\pm$ 0.2471 &  41.21089793234 $\pm$ 0.3646 & 19.2071 $\pm$ 0.0067 & 19.3055 $\pm$ 0.0434 & 18.6081 $\pm$ 0.0268\\ 
SDSS1004+4112            B  &  43  & 5 & 806853174703246208  & 151.14551708801 $\pm$ 0.2596 &  41.21187892534 $\pm$ 0.3323 & 19.2223 $\pm$ 0.0075 & 19.2207 $\pm$ 0.0422 & 18.7456 $\pm$ 0.0579\\ 
\hline
J1059+0622               B  &  32  & 4 & 3864688543049876352 & 164.86009927358 $\pm$20.6717 &	6.37436101583 $\pm$ 5.3105 & 17.9809 $\pm$ 0.0079 &                    &                    \\ 
J1059+0622               A  &  32  & 4 & 3864688543050432640 & 164.86015324956 $\pm$ 1.3350 &	6.37420550875 $\pm$ 1.4962 & 17.3898 $\pm$ 0.0238 & 17.2033 $\pm$ 0.0118 & 16.5761 $\pm$ 0.0121\\ 
\hline
HE1113-0641              B  &  01  & 4 & 3783971985705433984 & 169.09802843301 $\pm$ 6.0376 &  -6.96068252038 $\pm$ 1.8016 & 17.4980 $\pm$ 0.0266 & 16.7641 $\pm$ 0.0348 & 16.3431 $\pm$ 0.0227\\ 
HE1113-0641              A  &  01  & 4 & 3783971985705434112 & 169.09816765399 $\pm$ 7.0797 &  -6.96080328592 $\pm$ 5.1159 & 17.3972 $\pm$ 0.0137 & 16.8367 $\pm$ 0.0507 & 16.4789 $\pm$ 0.0240\\ 
\hline
PG1115+080               B  &  43  & 4 & 3817878828361980160 & 169.57015797740 $\pm$ 0.4371 &	7.76614341006 $\pm$ 0.2882 & 18.9413 $\pm$ 0.0109 &                    &                    \\ 
PG1115+080               A2 &  43  & 4 & 3817878828361980544 & 169.57066712476 $\pm$ 0.3243 &	7.76625027119 $\pm$ 0.3194 & 17.1502 $\pm$ 0.0146 & 16.5405 $\pm$ 0.0126 & 15.9208 $\pm$ 0.0158\\ 
PG1115+080               C  &  43  & 4 & 3817878828361980288 & 169.57025413596 $\pm$ 0.3157 &	7.76668844843 $\pm$ 0.1872 & 18.5837 $\pm$ 0.0070 & 18.5038 $\pm$ 0.0603 & 17.8989 $\pm$ 0.0371\\ 
PG1115+080               A1 &  43  & 4 & 3817878828362669568 & 169.57062805579 $\pm$24.4805 &	7.76612506243 $\pm$ 2.9684 & 17.1702 $\pm$ 0.0110 & 16.5525 $\pm$ 0.0223 & 15.9258 $\pm$ 0.0007\\ 
\hline
RXJ1131-1231             D  &  43  & 5 & 3586098513051815680 & 172.96404170566 $\pm$ 1.5657 & -12.53278020998 $\pm$ 0.7744 & 19.9911 $\pm$ 0.0199 &                    &                    \\ 
RXJ1131-1231             C  &  43  & 5 & 3586098513051815808 & 172.96475942625 $\pm$ 0.3308 & -12.53333404411 $\pm$ 0.2114 & 19.0189 $\pm$ 0.0141 &                    &                    \\ 
RXJ1131-1231             A  &  43  & 5 & 3586098513053631232 & 172.96492693884 $\pm$ 0.1390 & -12.53302324209 $\pm$ 0.0957 & 17.8773 $\pm$ 0.0149 & 17.1130 $\pm$ 0.0505 & 16.5723 $\pm$ 0.0442\\ 
RXJ1131-1231             B  &  43  & 5 & 3586098508760213248 & 172.96493514624 $\pm$ 0.2091 & -12.53269349597 $\pm$ 0.1275 & 18.3639 $\pm$ 0.0190 &                    &                    \\ 
\hline
2MASS J11344050-2103230  C  &  27  & 4 & 3541826024526317312 & 173.66852228570 $\pm$ 0.0768 & -21.05664892265 $\pm$ 0.0510 & 17.2709 $\pm$ 0.0045 & 17.4714 $\pm$ 0.0083 & 16.6770 $\pm$ 0.0627\\ 
2MASS J11344050-2103230  D  &  27  & 4 & 3541826024524572288 & 173.66873072407 $\pm$ 0.2741 & -21.05605458128 $\pm$ 0.1929 & 18.9371 $\pm$ 0.0061 &                    &                    \\ 
2MASS J11344050-2103230  A  &  27  & 4 & 3541826024524572544 & 173.66910193174 $\pm$ 0.0759 & -21.05643428362 $\pm$ 0.0506 & 17.1731 $\pm$ 0.0049 & 16.8891 $\pm$ 0.0443 & 16.2567 $\pm$ 0.0323\\ 
2MASS J11344050-2103230  B  &  27  & 4 & 3541826024526317568 & 173.66931901612 $\pm$ 0.0764 & -21.05594664876 $\pm$ 0.0502 & 17.1949 $\pm$ 0.0040 & 17.1767 $\pm$ 0.0759 & 16.4024 $\pm$ 0.0313\\ 
\hline
SDSS1138+0314            A  &  43  & 4 & 3800591477621812096 & 174.51563277165 $\pm$ 1.9150 &	3.24934233962 $\pm$ 0.8549 & 19.6834 $\pm$ 0.0104 & 19.0469 $\pm$ 0.0387 & 18.4332 $\pm$ 0.0394\\ 
\hline
SDSS J123903.30+444701.3 A  &  18* & 3 & 1541107486407607424 & 189.76374116747 $\pm$ 0.2836 &  44.78372677529 $\pm$ 0.2588 & 20.0381 $\pm$ 0.0036 & 20.3333 $\pm$ 0.0690 & 19.1140 $\pm$ 0.0335\\ 
SDSS J123903.30+444701.3 B  &  18* & 3 & 1541107490702900864 & 189.76436231033 $\pm$ 3.4623 &  44.78350620432 $\pm$ 4.3492 & 21.2902 $\pm$ 0.0260 &                    &                    \\ 
\hline
SDSSJ125107.57+293540.5  A  &  05  & 3 & 1464911708560136064 & 192.78156710722 $\pm$ 0.3323 &  29.59454644305 $\pm$ 0.2214 & 18.9284 $\pm$ 0.0123 & 18.5397 $\pm$ 0.0371 & 17.8071 $\pm$ 0.0252\\ 
SDSSJ125107.57+293540.5  B  &  05  & 3 & 1464911712855560960 & 192.78169178055 $\pm$33.1507 &  29.59464933771 $\pm$ 4.9416 & 20.1848 $\pm$ 0.0255 &                    &                    \\ 
\hline
2MASS J13102005-1714579  D  &  27  & 4 & 3511426761399393152 & 197.58291817500 $\pm$ 0.7841 & -17.25000870310 $\pm$ 0.6438 & 20.0227 $\pm$ 0.0126 & 19.8397 $\pm$ 0.1231 & 18.9536 $\pm$ 0.0674\\ 
2MASS J13102005-1714579  C  &  27  & 4 & 3511426761399771776 & 197.58326799853 $\pm$ 7.3493 & -17.24865451177 $\pm$ 3.8113 & 20.9737 $\pm$ 0.0417 &                    &                    \\ 
2MASS J13102005-1714579  B  &  27  & 4 & 3511426761399393280 & 197.58402962359 $\pm$ 0.7657 & -17.25007184536 $\pm$ 0.4609 & 19.7662 $\pm$ 0.0084 & 19.5670 $\pm$ 0.0685 & 18.5691 $\pm$ 0.0525\\ 
2MASS J13102005-1714579  A  &  27  & 4 & 3511426761399556352 & 197.58440002563 $\pm$ 0.6212 & -17.24934978580 $\pm$ 0.4847 & 19.8902 $\pm$ 0.0080 & 19.7893 $\pm$ 0.0433 & 19.0017 $\pm$ 0.0435\\ 
\hline
\\
SDSS J1330+1810          C  &  13  & 4 & 3746665350016537088 & 202.57739723361 $\pm$ 0.6044 &  18.17590345192 $\pm$ 0.3528 & 20.0199 $\pm$ 0.0097 &                    &                    \\ 
SDSS J1330+1810          AB &  13  & 4 & 3746665345725273472 & 202.57776001721 $\pm$ 0.8522 &  18.17557927148 $\pm$ 0.5600 & 18.9459 $\pm$ 0.0045 & 18.5218 $\pm$ 0.0191 & 17.6031 $\pm$ 0.0119\\ 
\hline
J1400+3134               B  &  13  & 3 & 1454504422981470976 & 210.05320429035 $\pm$ 0.4764 &  31.58171057446 $\pm$ 0.4428 & 20.0857 $\pm$ 0.0089 &                    &                    \\ 
J1400+3134               A  &  13  & 3 & 1454504418686043904 & 210.05354062959 $\pm$ 0.3133 &  31.58131960237 $\pm$ 0.2969 & 19.5755 $\pm$ 0.0077 & 19.6481 $\pm$ 0.0573 & 18.9430 $\pm$ 0.0411\\ 
\hline
H1413+117                C  &  43  & 4 & 1225461582386033536 & 213.94255533517 $\pm$ 9.6306 &  11.49554421221 $\pm$ 4.0495 & 17.5918 $\pm$ 0.0096 &                    &                    \\ 
H1413+117                A  &  43  & 4 & 1225461582386571008 & 213.94261368056 $\pm$ 1.7339 &  11.49530779513 $\pm$ 1.6421 & 17.3422 $\pm$ 0.0127 & 16.7056 $\pm$ 0.0129 & 15.9767 $\pm$ 0.0151\\ 
H1413+117                B  &  43  & 4 & 1225461582386033792 & 213.94281630572 $\pm$ 1.4809 &  11.49537124845 $\pm$ 1.7926 & 17.4443 $\pm$ 0.0161 & 17.0120 $\pm$ 0.1567 & 16.1650 $\pm$ 0.1682\\ 
\hline
B1422+231                C  &  43  & 4 & 1254357435158834560 & 216.15860664841 $\pm$ 0.0877 &  22.93328372661 $\pm$ 0.1121 & 16.8977 $\pm$ 0.0036 &                    &                    \\ 
B1422+231                B  &  43  & 4 & 1254357435159465088 & 216.15870787479 $\pm$ 0.9361 &  22.93349211808 $\pm$ 1.2745 & 16.6783 $\pm$ 0.0021 &                    &                    \\ 
B1422+231                A  &  43  & 4 & 1254357435158834688 & 216.15882701471 $\pm$ 0.5497 &  22.93357549720 $\pm$ 0.6104 & 16.2555 $\pm$ 0.0080 & 15.987 $\pm$ 0.017 & 14.962 $\pm$ 0.011\\ 
B1422+231                D  &  43  & 4 & 1254357435158834816 & 216.15899268510 $\pm$58.4203 &  22.93326929457 $\pm$34.3890 & 19.7110 $\pm$ 0.0149 &                    &                    \\ 
\hline
SDSS J1433+6007          C  &  38  & 4 & 1618050348046583680 & 218.34457378805 $\pm$ 0.7581 &  60.12083308456 $\pm$ 1.0440 & 20.2585 $\pm$ 0.0117 &                    &                    \\ 
SDSS J1433+6007          A  &  38  & 4 & 1618050348048027648 & 218.34499683862 $\pm$ 0.4977 &  60.12038164197 $\pm$ 0.4310 & 19.8749 $\pm$ 0.0076 & 19.6307 $\pm$ 0.0682 & 19.1394 $\pm$ 0.0539\\ 
SDSS J1433+6007          B  &  38  & 4 & 1618050554206457984 & 218.34499742143 $\pm$ 0.6334 &  60.12142437659 $\pm$ 0.5872 & 19.9861 $\pm$ 0.0094 & 19.8346 $\pm$ 0.0989 & 19.3034 $\pm$ 0.0479\\ 
\hline
J1606-2333               C  &  32  & 4 & 6242307087212540160 & 241.50098822162 $\pm$17.8936 & -23.55621065040 $\pm$13.6665 & 19.3273 $\pm$ 0.0097 &                    &        $\pm$	\\
J1606-2333               D  &  32  & 4 & 6242307087220282624 & 241.50089149589 $\pm$43.2108 & -23.55591724742 $\pm$ 7.8737 & 19.6110 $\pm$ 0.0163 &                    &        $\pm$	\\
J1606-2333               A  &  32  & 4 & 6242307087212540032 & 241.50122837627 $\pm$ 0.4723 & -23.55595910220 $\pm$ 0.1969 & 18.8525 $\pm$ 0.0053 & 18.2655 $\pm$ 0.0383 & 17.5597 $\pm$ 0.0275\\
J1606-2333               B  &  32  & 4 & 6242307087212539776 & 241.50073780328 $\pm$ 1.3337 & -23.55612324162 $\pm$ 0.3274 & 18.9721 $\pm$ 0.0115 & 18.3351 $\pm$ 0.0536 & 17.5623 $\pm$ 0.0797\\
\hline
J1721+8842               C  &  32  & 4 & 1729026461820249728 & 260.41335642070 $\pm$ 0.6103 &  88.70589729174 $\pm$ 0.4895 & 19.9265 $\pm$ 0.0072 &                    &        $\pm$	   \\ 
J1721+8842               B  &  32  & 4 & 1729026466114871424 & 260.43780732365 $\pm$ 0.2382 &  88.70660390446 $\pm$ 0.2481 & 19.0310 $\pm$ 0.0050 &                    &        $\pm$	   \\ 
J1721+8842               D  &  32  & 4 & 1729026466114871296 & 260.44440309955 $\pm$ 3.4388 &  88.70521994925 $\pm$ 1.0992 & 20.6595 $\pm$ 0.0174 &                    &        $\pm$	   \\ 
J1721+8842               A  &  32  & 4 & 1729026466116588544 & 260.45419370846 $\pm$ 0.1206 &  88.70621468902 $\pm$ 0.1300 & 18.1755 $\pm$ 0.0053 & 18.3643 $\pm$ 0.0387 & 17.1173 $\pm$ 0.0174\\ 
J1721+8842              ?*  &  32  & 4 & 1729026461820390400 & 260.45126749084 $\pm$ 0.1535 &  88.70556237840 $\pm$ 0.1721 & 18.3287 $\pm$ 0.0084 & 18.4834 $\pm$ 0.0811 & 17.3640 $\pm$ 0.0401\\	
\hline
WFI2026-4536             B  &  43  & 4 & 6675746940384195456 & 306.54348562606 $\pm$ 0.3228 & -45.60718787581 $\pm$ 0.2604 & 18.6226 $\pm$ 0.0079 &                    &                    \\ 
WFI2026-4536             A  &  43  & 4 & 6675746940384195200 & 306.54363561758 $\pm$ 3.0726 & -45.60752504082 $\pm$ 1.6543 & 17.1982 $\pm$ 0.0132 & 16.8025 $\pm$ 0.0290 & 16.2132 $\pm$ 0.0125\\ 
\hline
WFI2033-4723             A1 &  43  & 4 & 6674418764699092736 & 308.42535000446 $\pm$ 0.2748 & -47.39537499737 $\pm$ 0.2933 & 18.0677 $\pm$ 0.0071 & 17.7333 $\pm$ 0.0173 & 17.0862 $\pm$ 0.0127\\ 
WFI2033-4723             C  &  43  & 4 & 6674418764698005248 & 308.42538394120 $\pm$ 0.2748 & -47.39580256979 $\pm$ 0.2726 & 19.2447 $\pm$ 0.0092 &                    &        $\pm$	   \\ 
WFI2033-4723             A2 &  43  & 4 & 6674418764698005376 & 308.42564273060 $\pm$ 0.1850 & -47.39534323862 $\pm$ 0.2475 & 18.8196 $\pm$ 0.0050 &                    &        $\pm$	   \\ 
WFI2033-4723             B  &  43  & 4 & 6674418764698004992 & 308.42625134682 $\pm$ 0.1890 & -47.39572539107 $\pm$ 0.1672 & 18.8262 $\pm$ 0.0023 & 18.7933 $\pm$ 0.0767 & 18.1817 $\pm$ 0.0680\\ 
\hline
WGD2038-4008            C*  &  35  & 4 & 6681326549578891648 & 309.51078424930 $\pm$ 0.5758 & -40.13693605210 $\pm$ 0.4304 & 19.9010 $\pm$ 0.0055 &                    &                    \\ 
WGD2038-4008            A*  &  35  & 4 & 6681326549580116864 & 309.51107237693 $\pm$ 0.5785 & -40.13740161713 $\pm$ 0.5994 & 19.6071 $\pm$ 0.0095 & 18.9346 $\pm$ 0.1256 & 17.3375 $\pm$ 0.0376\\ 
WGD2038-4008            D*  &  35  & 4 & 6681326549578891392 & 309.51157347188 $\pm$ 2.1359 & -40.13683014244 $\pm$ 1.7719 & 20.2563 $\pm$ 0.0144 &                    &                    \\ 
WGD2038-4008            B*  &  35  & 4 & 6681326549578891520 & 309.51162267018 $\pm$ 0.5164 & -40.13740966159 $\pm$ 0.4163 & 19.6480 $\pm$ 0.0053 &                    &                    \\ 
\hline
PS1 J205143-111444       A  &  40* & 4 & 6901910950299842688 & 312.93087874749 $\pm$ 0.5607 & -11.24562339171 $\pm$ 0.3446 & 19.6632 $\pm$ 0.0052 & 19.2664 $\pm$ 0.0422 & 18.5697 $\pm$ 0.0377\\ 
PS1 J205143-111444       B  &  40* & 4 & 6901910950301201024 & 312.93122111259 $\pm$ 0.5220 & -11.24509460838 $\pm$ 0.2868 & 19.9321 $\pm$ 0.0055 & 19.9348 $\pm$ 0.1826 & 18.8598 $\pm$ 0.0397\\ 
PS1 J205143-111444       C  &  40* & 4 & 6901910950299842560 & 312.93160192750 $\pm$ 1.5900 & -11.24505252971 $\pm$ 1.4157 & 20.7426 $\pm$ 0.0163 &                    &                    \\ 
\hline
WGD2141-4629             B  &  35* & 3 & 6563636302411599104 & 325.45348370389 $\pm$ 1.0225 & -46.49631823151 $\pm$ 0.6141 & 20.3767 $\pm$ 0.0101 &                    &                    \\ 
WGD2141-4629             A  &  35* & 3 & 6563636302410205824 & 325.45360046795 $\pm$ 0.5330 & -46.49607609394 $\pm$ 0.5227 & 19.8817 $\pm$ 0.0093 & 19.8634 $\pm$ 0.0637 & 18.8872 $\pm$ 0.0409\\ 
WGD2141-4629             C  &  35* & 3 & 6563636302410673152 & 325.45423455272 $\pm$ 0.9792 & -46.49553407828 $\pm$ 1.4042 & 20.8928 $\pm$ 0.0118 & 21.6045 $\pm$ 0.2997 & 20.4365 $\pm$ 0.1651\\ 
\hline
Q2237+030                B  &  43  & 4 & 2704542594213521664 & 340.12576696218 $\pm$ 0.5140 &	3.35881124314 $\pm$ 0.4528 & 17.6672 $\pm$ 0.0097 &                    &                    \\ 
Q2237+030                A  &  43  & 4 & 2704542589921820288 & 340.12595348968 $\pm$ 0.3627 &	3.35834194908 $\pm$ 0.2799 & 16.7356 $\pm$ 0.0109 & 16.2755 $\pm$ 0.0423 & 15.1978 $\pm$ 0.0306\\ 
\hline
\hline
\multicolumn{9}{l}{\tablefoot{1. WGD2038-4008* : the images (A, B, C, D) have been attributed following increasing $G$ mag, and not following \citet{2017Agnelloa} since even considering the photometric data from the aforementioned paper this would be the decreasing flux order.}}\\
\multicolumn{9}{l}{\tablefoot{2. J0408-5354* : the images (A, B, C) have been attributed following increasing $G$ mag, and not following \citet{2017Lin}.}}\\
\multicolumn{9}{l}{
\tablebib{(1)~\cite{2008Blackburn}, (2)~\cite{2008Jackson}, (3)~\cite{2009Ghosh}, (4)~\cite{2010Inada}, (5)~\cite{2012Inada}, (6)~\cite{2012Jackson}, (7)~\cite{2014Inada}, (8)~\cite{2015Agnello},(9)~\cite{2016Limousin}, (10)~\cite{2016More}, (11)~\cite{2016Inoue},(12)~\cite{2016Aravena}, (13)~\cite{2016Rusu}, (14)~\cite{2016Goicoechea}, (15)~\cite{2016Leethochawalit}, (16)~\cite{2016Nayyeri}, (17)~\cite{2016Parry}, (18)~\cite{2016Sergeyev}, (19)~\cite{2016Shu}, (20)~\cite{2017More}, (21)~\cite{2017Agnelloa}, (22)~\cite{2017Agnellob}, (23)~\cite{2017Lin}, (24)~\cite{2018Ostrovski}, (25)~\cite{2018Kostrzewa-Rutkowska}, (26)~\cite{2018Lemon}, (27)~\cite{2018Lucey}, (28)~\cite{2018Williams}, (29)~\cite{2018Kostrzewa-Rutkowska}, (30)~\cite{2016Bordoloi}, (31)~\cite{2002Wisotzki}, (32)~\cite{2018Lemon}, (33)~\cite{2017Berghea}, (34)~\cite{2017More}, (35)~\cite{2017Agnelloa}, (36)~\cite{2013Dahle}, (37)~\cite{2018Schechter}, (38)~\cite{2017Agnellob}, (39)~\cite{2017Agnelloa}, (40)~\cite{2018Rusu}, (41)~\cite{2017Meyer}, (42)~\cite{2017Schechter}, (43)~\cite{1999Castles}, (44)~\cite{2009Anguita}}
}
\end{longtable}
\end{landscape}
}

\begin{appendix}
\section{Gaia DR2 finding charts of known and confirmed quadruply-imaged quasars}

\begin{figure*}[h!]
	\centering
\includegraphics[width=0.31\textwidth]{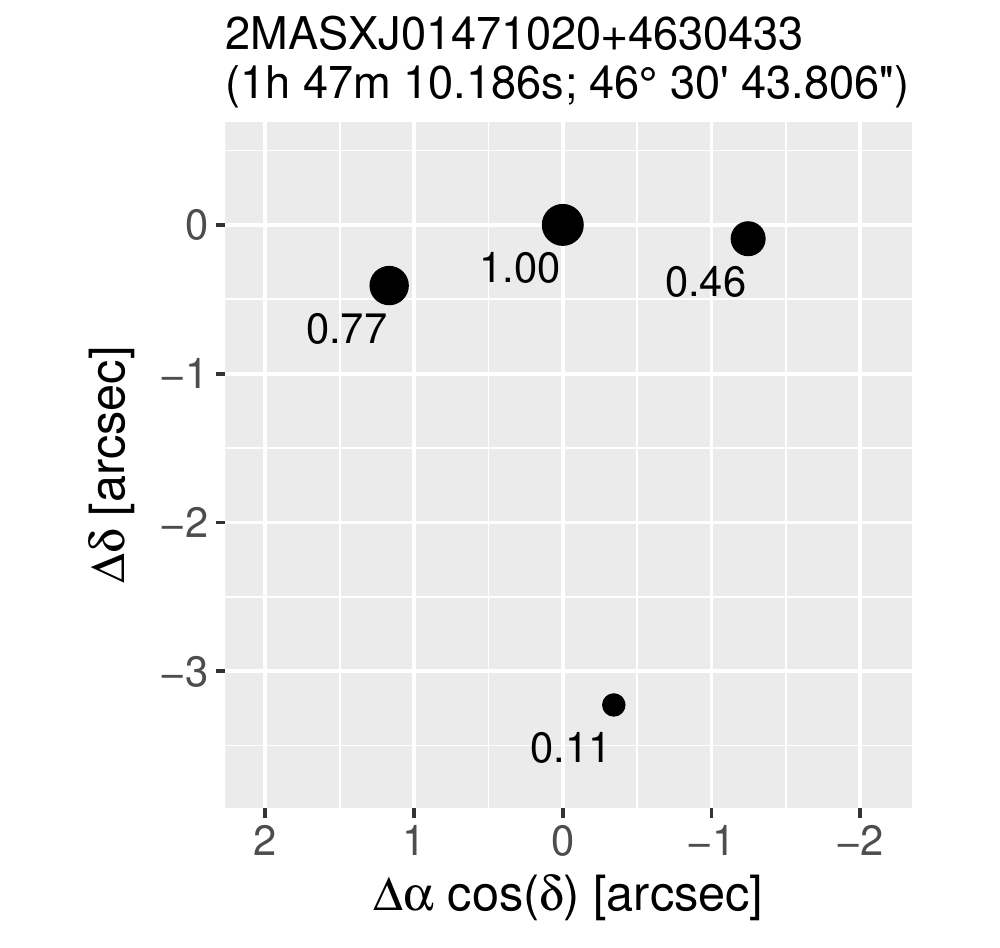}
	\centering
\includegraphics[width=0.31\textwidth]{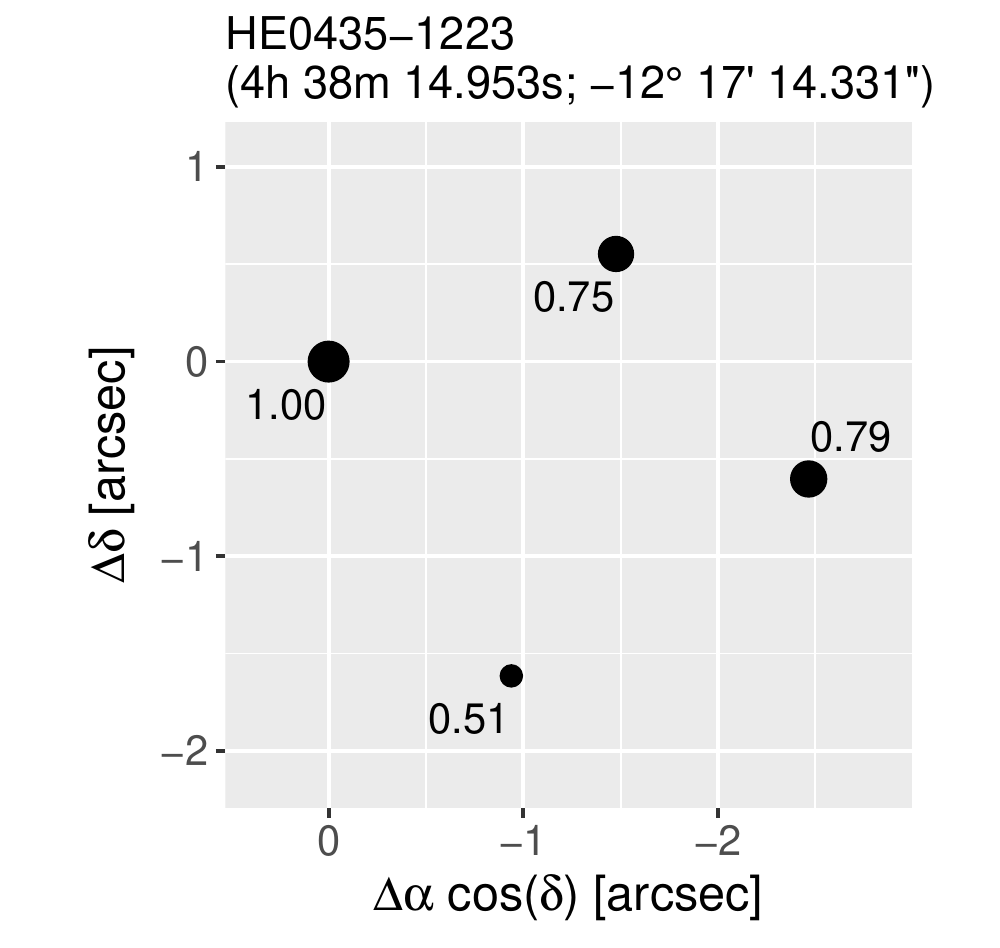}
	\centering
\includegraphics[width=0.31\textwidth]{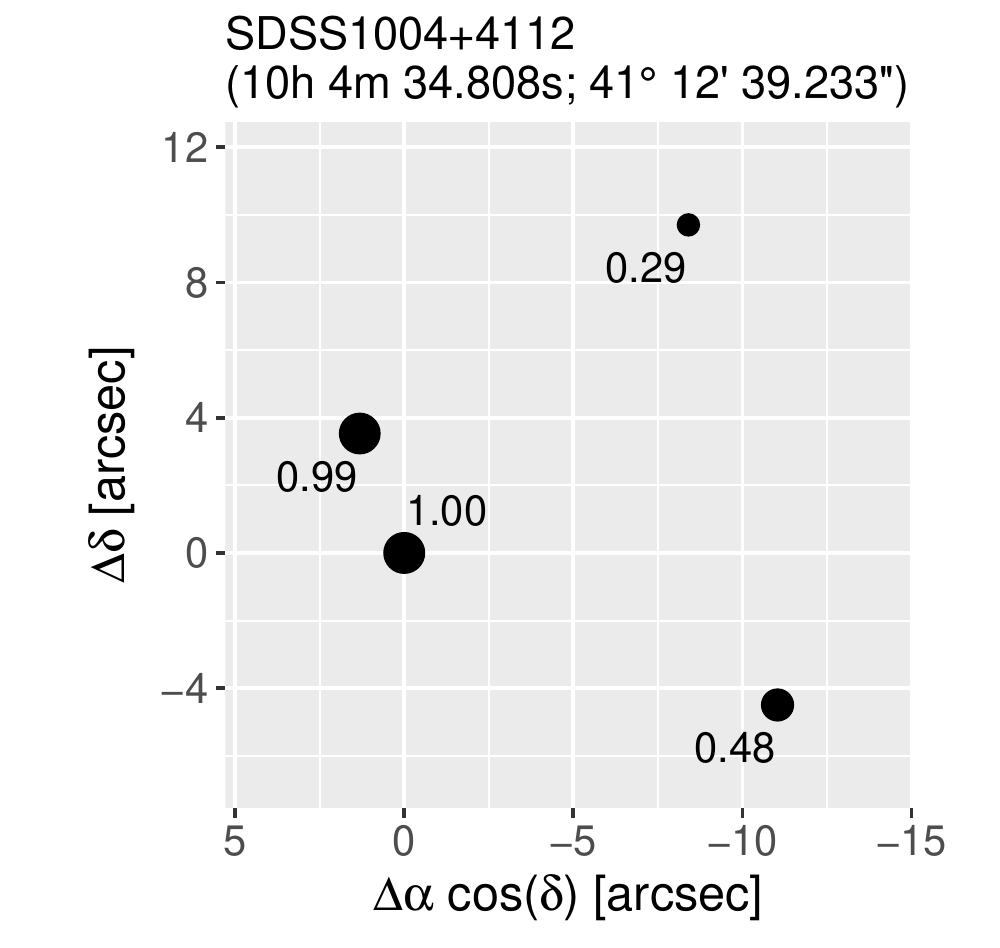}

	\centering
\includegraphics[width=0.31\textwidth]{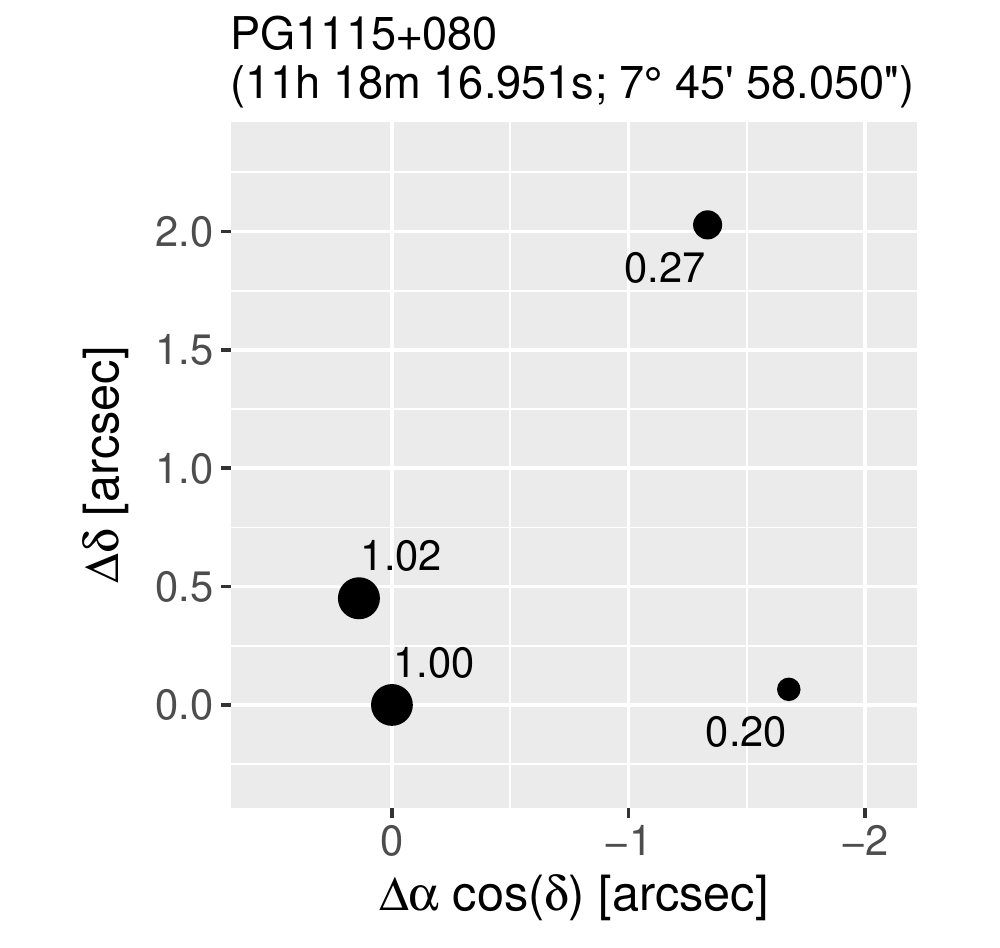}
    \centering
\includegraphics[width=0.31\textwidth]{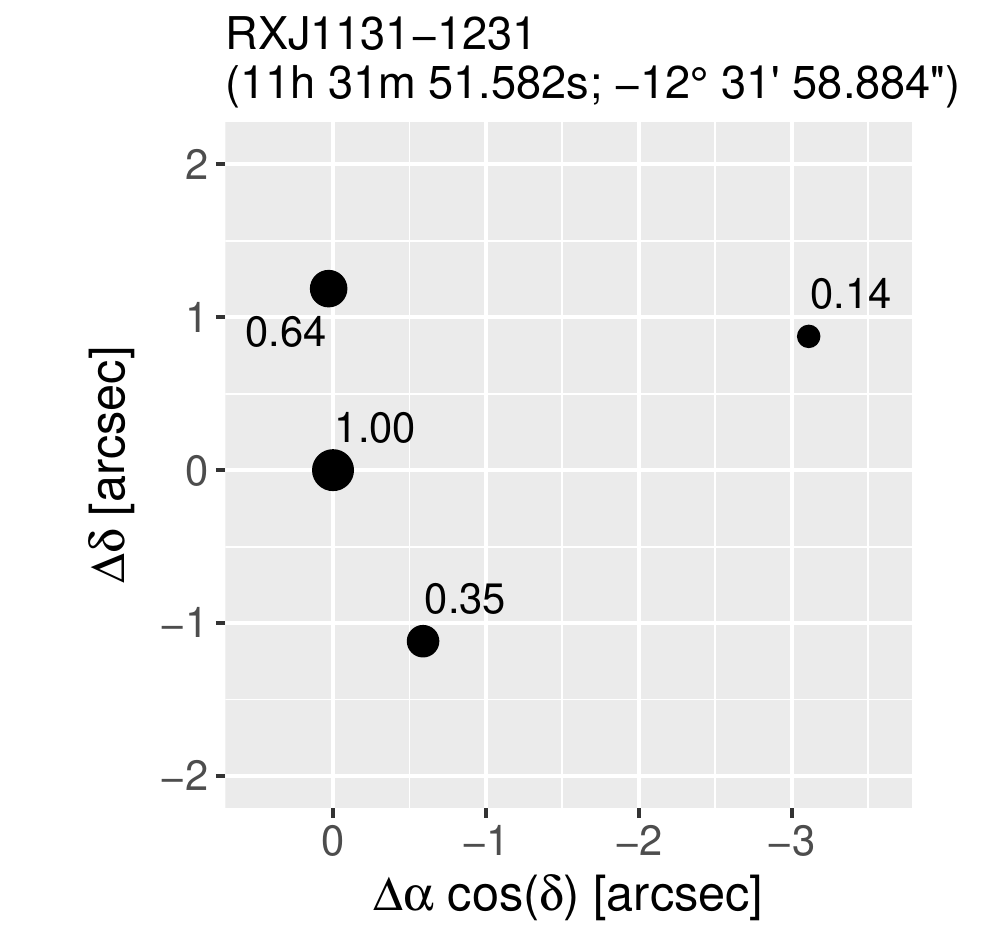}
	\centering
\includegraphics[width=0.31\textwidth]{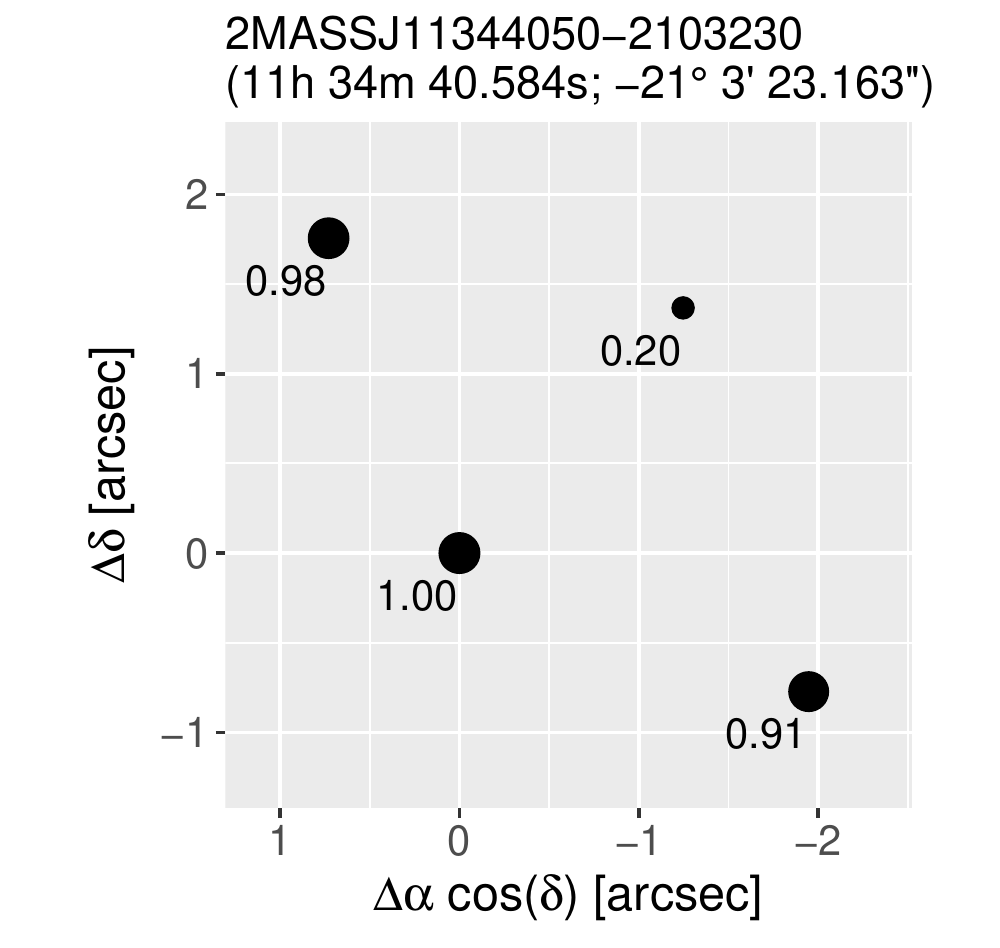}

	\centering
\includegraphics[width=0.31\textwidth]{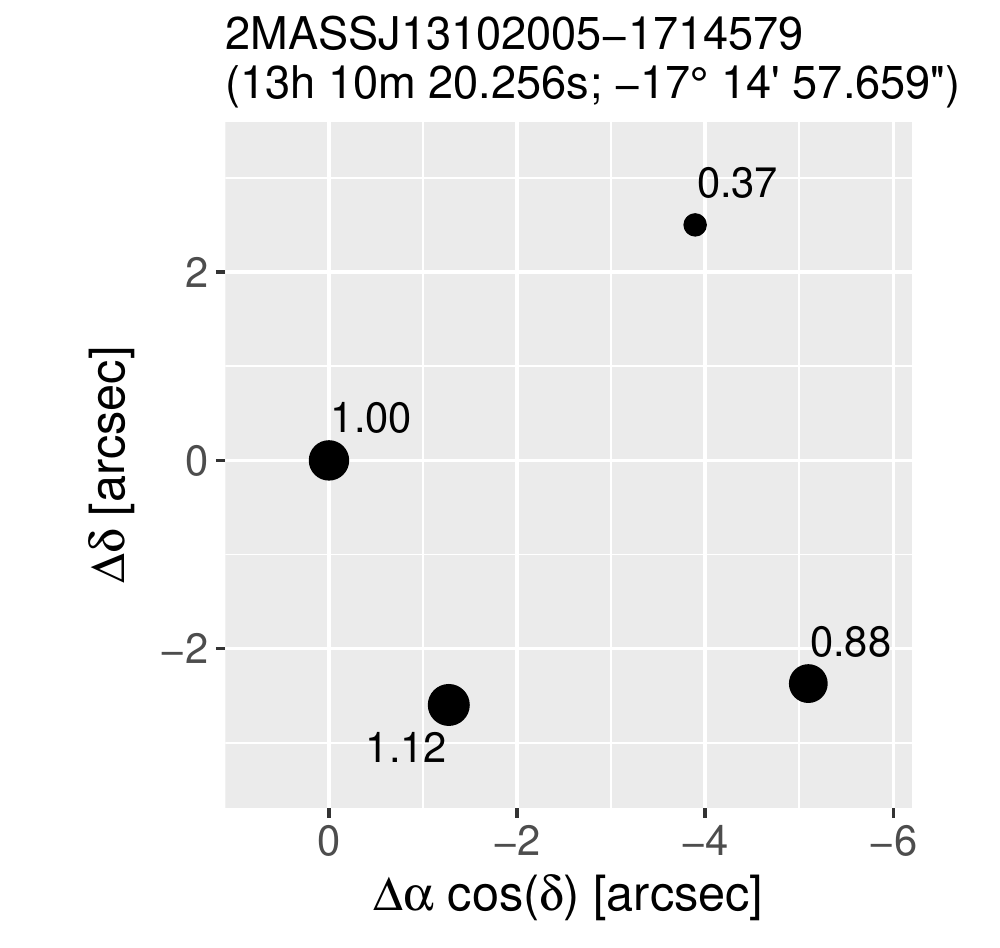}
	\centering
\includegraphics[width=0.31\textwidth]{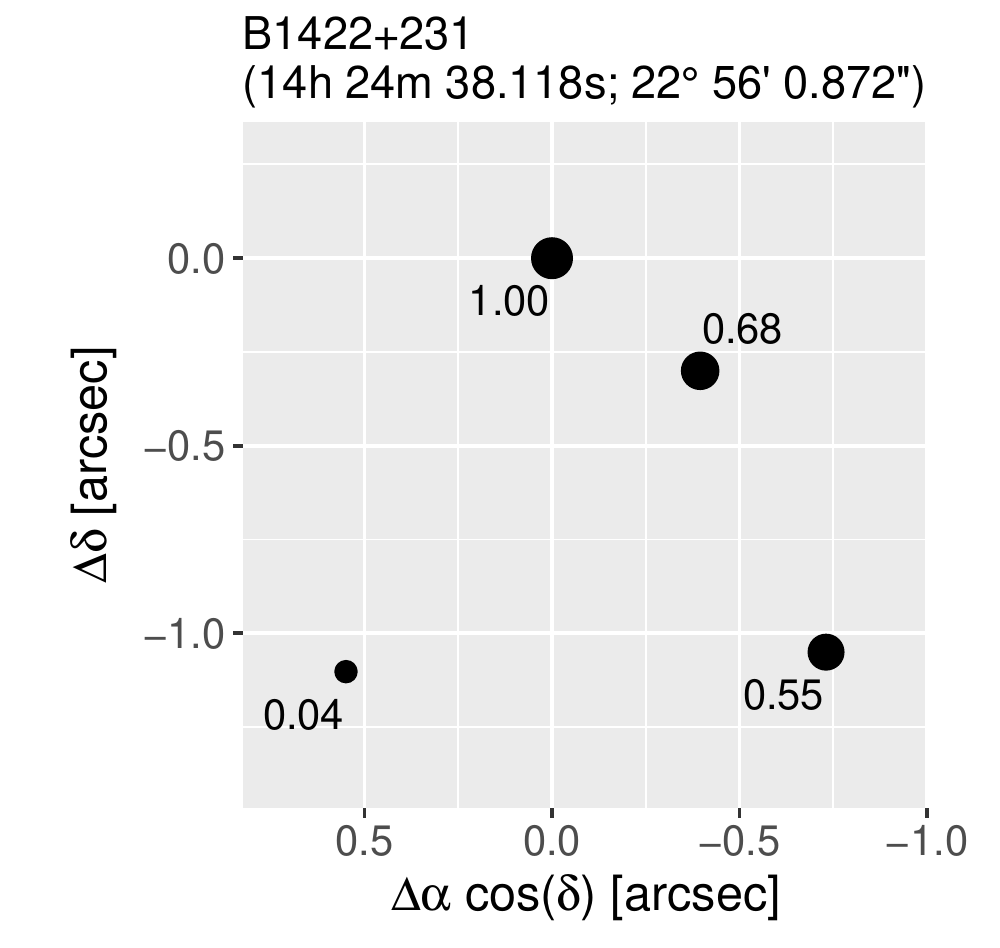}
	\centering
\includegraphics[width=0.31\textwidth]{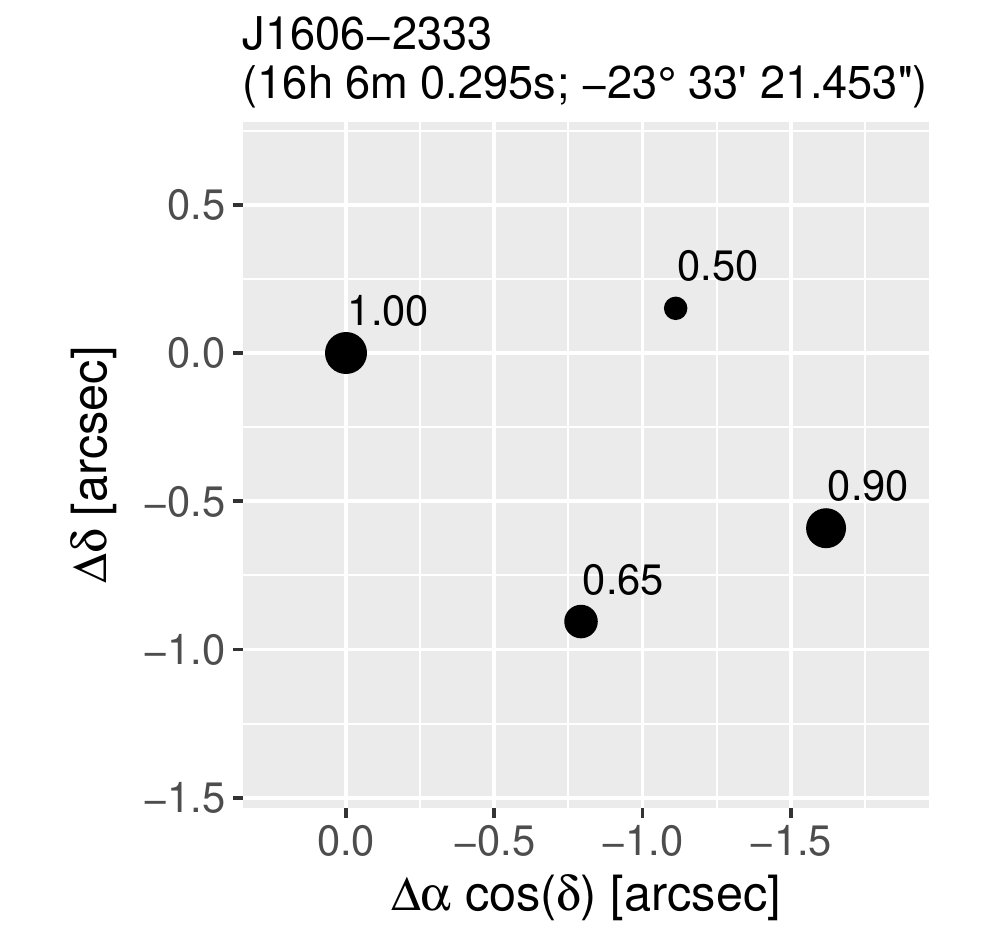}

	\centering
\includegraphics[width=0.31\textwidth]{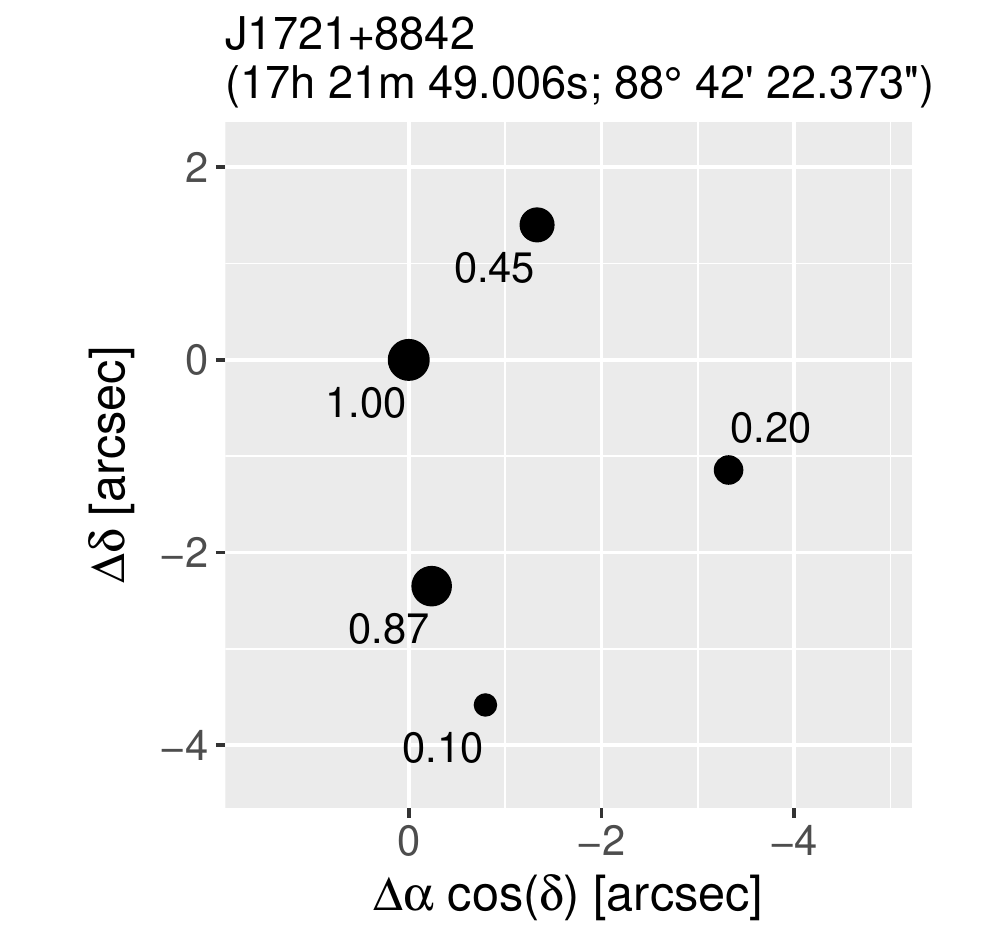}
	\centering
\includegraphics[width=0.31\textwidth]{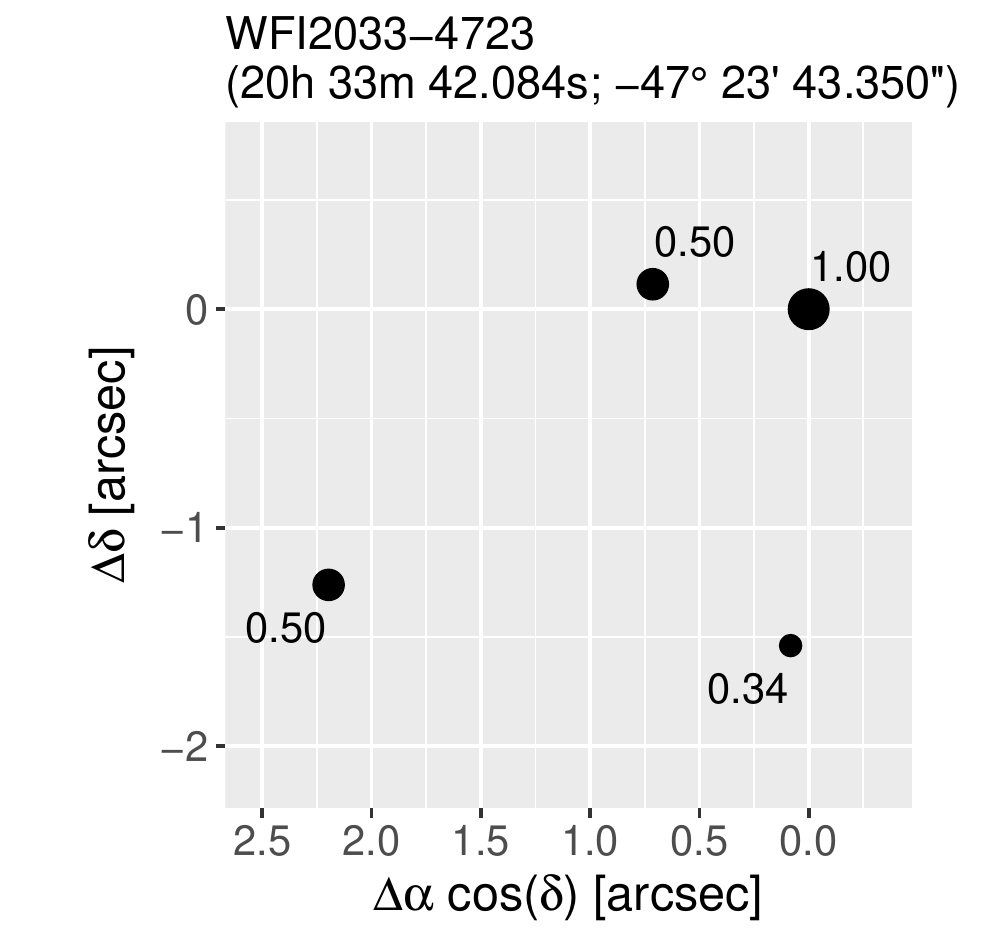}
	\centering
\includegraphics[width=0.31\textwidth]{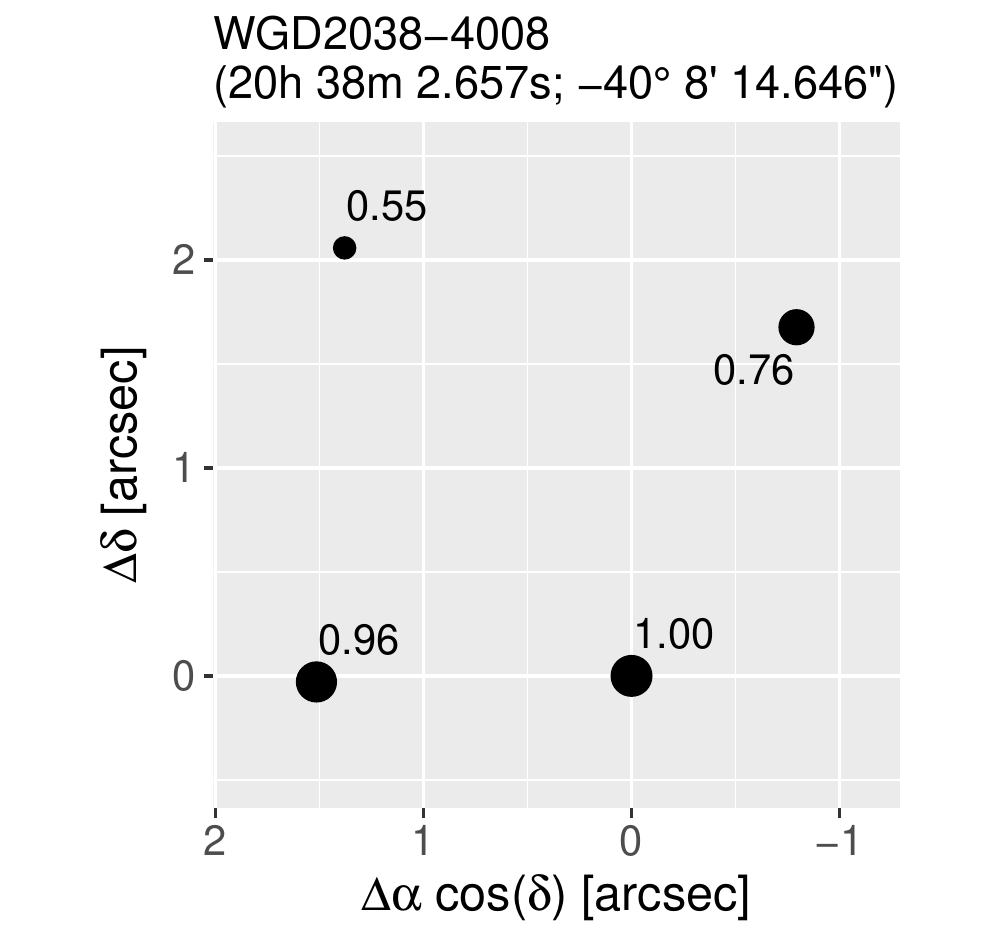}
\caption{Finding charts for the 12 previously known multiply-imaged quasars with four counterparts in the {\it Gaia} Data Release 2. {\it Gaia} DR2 astrometry relative to the brightest image in the system discovery passband (image ``A'') is indicated by black points, except for WGD2038-4008, which is ordered by Gaia $G$-band fluxes (see footnote in Table \ref{knownlensDR2_34}). The numbers near each image, and the image size, indicate the flux ratios computed from {\it Gaia} DR2 data. North is up, East is left. 
    \label{charts}}
\end{figure*}

\end{appendix}

\end{document}